\documentclass[acmlarge]{acmart}
\usepackage{graphicx}
\usepackage{subfigure}
\usepackage{booktabs} % For formal tables
\usepackage{enumitem}
\usepackage{amsmath}
\usepackage{url}
\usepackage{esvect}
\usepackage{varwidth}
\usepackage[ruled,linesnumbered]{algorithm2e} % For algorithms
\usepackage{multirow}
\usepackage{siunitx} 
\usepackage{color, soul}
\soulregister\cite7
\soulregister\ref7
\soulregister\pageref7

\usepackage[ruled]{algorithm2e} % For algorithms

\SetAlFnt{\small}
\SetAlCapFnt{\small}
\SetAlCapNameFnt{\small}
\SetAlCapHSkip{0pt}
\IncMargin{-\parindent}
\graphicspath{{figures/}}

\acmJournal{IMWUT}
\acmVolume{2}
\acmNumber{3}
\acmArticle{xx}
\acmYear{2018}
\acmMonth{9}
\acmArticleSeq{1}

\setcopyright{acmcopyright}
\acmDOI{00.0000/0000000}

\received{February 2018}
\received{May 2018}
\received[accepted]{September 2018}

\begin{document}
% Title portion
\title{Authenticating On-Body Backscatter by Exploiting Propagation Signatures} 
% \titlenote{We can add a note to the title}
\author{Zhiqing Luo}
\email{zhiqing_luo@hust.edu.cn}
\author{Wei Wang}
\authornote{This is the corresponding author}
\email{weiwangw@hust.edu.cn}
\author{Jiang Xiao}
\email{jiangxiao@hust.edu.cn}
\affiliation{%
	\institution{Huazhong University of Science and Technology}
	\streetaddress{1037 Luoyu Road}
	\city{Wuhan}
	\state{Hubei}
	\country{China}}

\author{Qianyi Huang}
%\authornote{This is the corresponding author}
\email{qhuangaa@connect.ust.hk}
\affiliation{%
	\institution{Hong Kong University of Science and Technology}
	\streetaddress{Clear Water Bay, Kowloon}
	\city{Hong Kong}
	\state{Hong Kong}
	\country{China}}
\author{Tao jiang}
\email{taojiang@hust.edu.cn}
%\author{XX}
%%\authornote{This is the corresponding author}
%\email{XX}
\affiliation{%
  \institution{Huazhong University of Science and Technology}
  \streetaddress{1037 Luoyu Road}
  \city{Wuhan}
  \state{Hubei}
  \country{China}}
\author{Qian Zhang}
%\authornote{This is the corresponding author}
\email{qianzh@cse.ust.hk}
\affiliation{%
	\institution{Hong Kong University of Science and Technology}
	\streetaddress{Clear Water Bay, Kowloon}
	\city{Hong Kong}
	\state{Hong Kong}
	\country{China}}

%!TEX root=main.tex

\begin{abstract}
    The vision of battery-free communication has made backscatter a compelling technology for on-body wearable and implantable devices. Recent advances have facilitated the communication between backscatter tags and on-body smart devices. These studies have focused on the communication dimension, while the security dimension remains vulnerable. It has been demonstrated that wireless connectivity can be exploited to send unauthorized commands or fake messages that result in device malfunctioning. The key challenge in defending these attacks stems from the minimalist design in backscatter. Thus, in this paper, we explore the feasibility of authenticating an on-body backscatter tag without modifying its signal or protocol. We present~\textit{SecureScatter}, a physical-layer solution that delegates the security of backscatter to an on-body smart device. To this end, we profile the on-body propagation paths of backscatter links, and construct highly sensitive propagation signatures to identify on-body backscatter links. We implement our design in a software radio and evaluate it with different backscatter tags that work at 2.4~GHz and 900~MHz. Results show that our system can identify on-body devices at 93.23\% average true positive rate and 3.18\% average false positive rate.

\end{abstract}





% DO NOT use this command unless you want to change
% the default behavior
% \authorsaddresses{Authors' addresses: G.~Zhou, Computer Science
%   Department, College of William and Mary, 104 Jameson Rd,
%   Williamsburg, PA 23185, US, \path{gzhou@wm.edu}; V.~B\'eranger,
%   Inria Paris-Rocquencourt, Rocquencourt, France; A.~Patel, Rajiv
%   Gandhi University, Rono-Hills, Doimukh, Arunachal Pradesh, India;
%   H.~Chan, Tsinghua University, 30 Shuangqing Rd, Haidian Qu, Beijing
%   Shi, China; T.~Yan, Eaton Innovation Center, Prague, Czech Republic;
%   T.~He, C.~Huang, J.~A.~Stankovic University of Virginia, School of
%   Engineering Charlottesville, VA 22903, USA; T. F. Abdelzaher,
%   (Current address) NASA Ames Research Center, Moffett Field,
%   California 94035.}

\maketitle

\renewcommand{\shortauthors}{Z. Luo et al.}

\section{Introduction} \label{sec:introduction}
%1. backscatter features. backscatter has been envisioned to be applied to on-body wearables and medical devices.
The ultra low-power nature of backscatter communication makes it a promising technology to enable battery-free communication for on-body wearable and implantable devices with tiny energy budgets. These wearables and implantables include smart contact lens~\cite{pandey2010fully,yao2012contact}, glucose sensor~\cite{liao20123}, implantable neural device~\cite{yin2012100}, as well as flexible wearables~\cite{pantelopoulos2010survey,wang2017sampleless}. The battery-free communication enabled by backscatter allows on-body sensor data to be continuously streamed to nearby data hub devices such as a smartphone or smart watch~\cite{interscatter,onbodyscatter}. The ultra low-power nature combined with the simple hardware components of backscatter-based sensors perfectly matches the need of miniature on-body devices.

%2. they have focused on communication, while security issue is missed. it is very severe in IMD. must be tackled. authentication is one important aspect.

\begin{figure}[t]
    \center 
    \subfigure[Signal from an on-body tag.]
    {\includegraphics[width=3.5in]{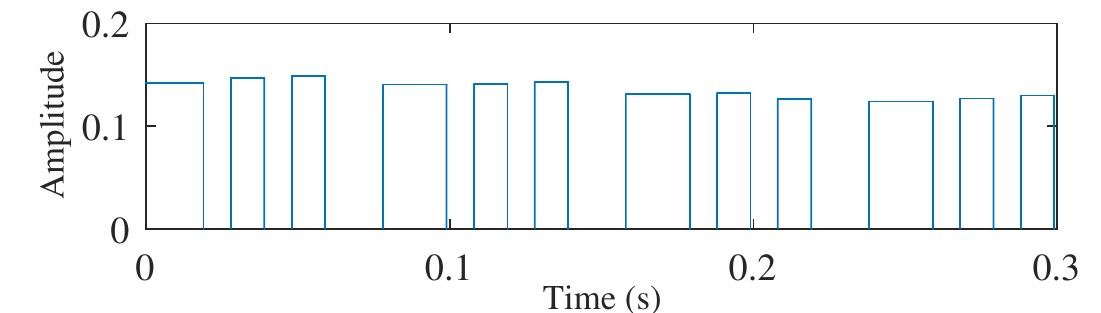}}
    \subfigure[Signal from an attacker.]
    {\includegraphics[width=3.5in]{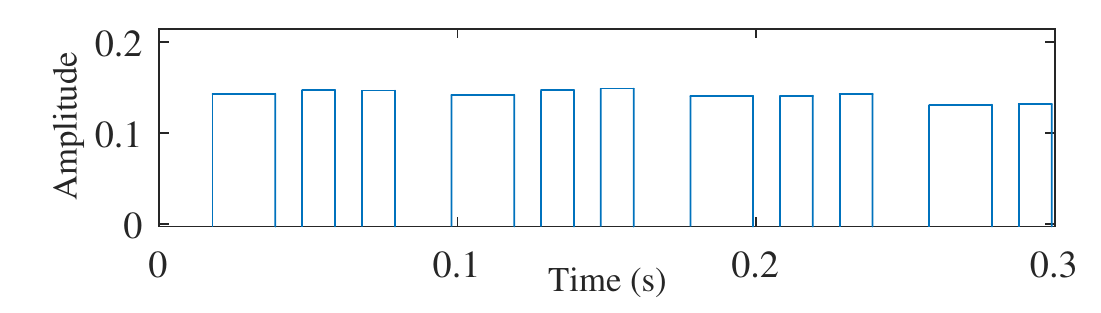}}
    %    \subfigure[Raw off-body RSS.]
    %    {\hspace{0.35cm}\includegraphics[width=3.0in]{attacker.pdf}}
    
    %    \subfigure[Off-body RSS after fluctuation removal.]
    %    {\hspace{0.35cm}\includegraphics[width=1.65in]{figs//new/ica_plot/off_removal}}
    \caption{An active attacker emulates the reflection patterns of backscatter to perform a spoofing attack. The results show the decoded waveforms at the receiver.}
    \label{fig:tag} \vspace{-0.3cm}
\end{figure}

The latest advances in backscatter communications~\cite{interscatter,onbodyscatter} have made it practical to enable an on-body backscatter to reflect transmissions of an on-body link, e.g., a link between a smart watch and a smartphone~\cite{interscatter}. These studies have focused on the communication dimension, while the security dimension of backscatter remains unexplored. The minimalist design of backscatter appears to be a double-edged sword: it opens up a range of possibilities for ultra low-power communications but makes the communication vulnerable to malicious attacks. Unfortunately, these wearables and implantables measure users' daily activities, vital signs, and even deliver timely treatment, which are extremely sensitive. Recent works have shown that wireless connectivity can be compromised to send unauthorized commands to make implantable devices malfunctioning, even make it deliver an electric shock to users~\cite{gollakota2011they,halperin2008security}. In addition, backscatter signals can be easily faked. Our experiment shows that a transmitter referred to as an active attacker from afar can imitate the behaviors of backscatter to send emulated reflected signals, as illustrated in Fig.~\ref{fig:tag}. We observe that without a dedicated authentication mechanism, the receiver cannot discriminate the valid backscatter signal from the signal from an attacker.

%3. traditional security schemes does not work as backscatter is minimalist design. cannot compute or perform complex protocols.

To make backscatter a practical low-power solution for on-body devices, device authentication is an essential security feature to prevent unauthorized access. With an authentication mechanism, devices can verify the identity of the sender to avoid logging fake data or commands sent by malicious devices. In other systems, such as Wi-Fi and Bluetooth, cryptographic methods or PIN codes are utilized to prevent unauthorized access. However, incorporating these methods into backscatter tags is difficult due to the minimalist design of backscatter. Backscatter circuits drop most of the digital modules to reduce energy consumption while leaving merely indispensable components for signal reflection and demodulation. Thus, backscatter-enabled devices cannot afford complex computation overhead imposed by these security mechanisms.

%4. our observation is that on-body backscatter normally communicate with another on-body smart device which can be leverage to delegate backscatter for authentication. and the on-body propagation exhibits distinct patterns that can be leveraged for authentication.
\begin{figure}[htbp]
	\centering
	\includegraphics[width=0.42\linewidth]{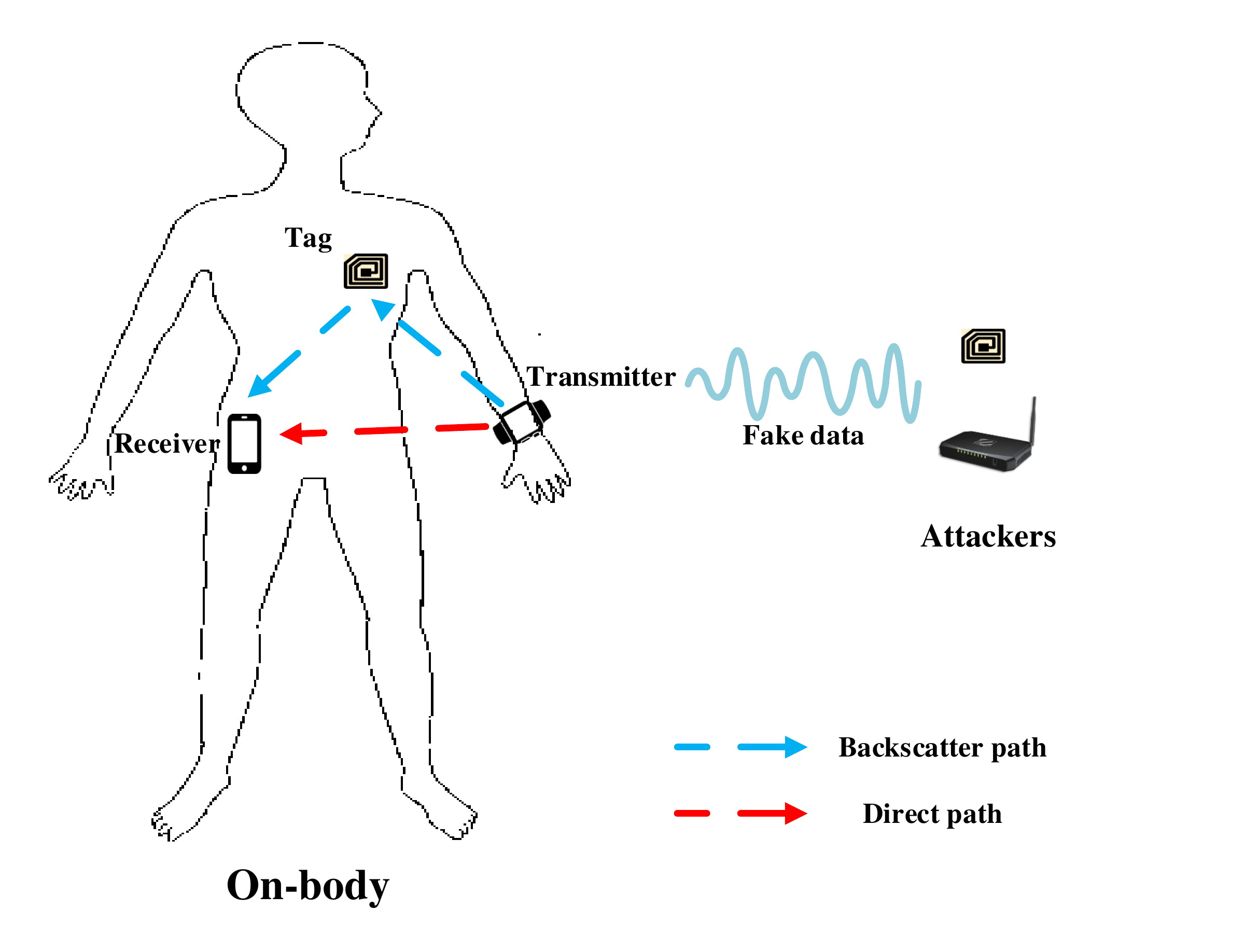} 
	\caption{Illustration of on-body backscatter and off-body attack. On-body transmitter communicates with the receiver, and the on-body tag establishes the link and transmit the data to the receiver by backscattering the direct path. An off-body active attacker transmits the fake data to the receiver by emulating the on-body tag}
	\label{fig:model}%\vspace{-0.2cm}
\end{figure}
This paper takes the first step to explore the feasibility of authenticating on-body backscatter devices without modifying the signal and protocol. We leverage on-body smart devices that the backscatter device communicates with to delegate the backscatter device to perform all authentication operations. We present a design called \textit{SecureScatter}, a physical-layer solution that exploits the distinct on-body backscatter propagation to verify a sender's identity. Our key insight is that the on-body propagation is dominated by a special type of waves, i.e., creeping waves diffracted from human tissue and trapped along the body's surface~\cite{ryckaert2004channel,mcnamara1990uniform,pethig1987dielectric}. Different from off-body transmissions, the reflection paths of on-body backscatter exhibit distinct patterns in that they are extremely sensitive to device movement, body movements, and antenna directions. These features motivate us to exploit the signatures lying in the on-body backscatter propagation to differentiate the genuine on-body backscatter devices from off-body backscatter devices referred to as tag attackers or active attackers emulating the behaviors of backscatter from afar.

Realizing the above idea entails the following challenges.

\textit{1) How to exploit the on-body backscatter propagation signatures to authenticate on-body tags without any hardware changes or extra hardware?} The minimalist design of backscatter makes it impossible to add extra components to the backscatter devices. In addition, an antenna array~\cite{xiong2013securearray} that profiles fine-grained information of received signals cannot be applied to our case as on-body smart devices such as smartphone and smart watch cannot afford antenna arrays. In order to surmount this obstacle, we introduce a simple movement-based authentication method that utilizes a simple transmitter movement to assist smart devices to capture propagation signatures using a single antenna. The transmitter movement excites the variations in the backscatter link as well as the link between on-body smart devices. Such variations contain salient on-body propagation signatures related to antenna polarization and movement correlations among on-body propagation paths, which can be extracted to identify on-body backscatter tags.

\textit{2) How to differentiate whether the variations are caused by dynamic body movement, authentication movements or powerful active attackers?} On-body propagation is easily affected by human movements, which makes it non-trivial to differentiate the variations caused by authentication movement and the unintentional body movements. In order to overcome this predicament, SecureScatter firstly investigates the variation patterns of backscatter signal and the direct path that the signals directly transmit from the transmitter to the receiver. Accordingly, SecureScatter is able to extract the variation features caused by authentication or body movement effects. Based on the different variation scales, SecureScatter can remove the effects of dynamic body movement. Besides, a powerful active attacker who transmits varying powers can also imitate these variation scales and attack the on-body tag. To solve this problem, SecureScatter observes that the variations between on-body backscatter path and direct path have high correlations, which can be used to defend the powerful attackers.

\textbf{Summary of results.} We evaluate SecureScatter using different backscatter tags that work at 2.4~GHz and 900~MHz against both active and tag attackers. On the whole, SecureScatter achieves an average true positive (TP) rate of $93.23\%$ and false positive (FP) rate of $3.18\%$ in various static and dynamic environments. Additionally, SecureScatter yields an acceptable latency of 150~ms.

\textbf{Contributions.} In summary, we make the following contributions. First, we show that the propagation signatures of on-body backscatter can be leveraged to authenticate on-body backscatter. Second, we develop SecureScatter, an on-body backscatter authentication framework that requires no extra process or hardware at the backscatter tag, and thus aligns with the minimalist design of backscatter. Finally, we test SecureScatter under different static and dynamic environments, and the results show that the effectiveness against both active and tag attackers.
%\begin{figure}[t]
 %   \center
 %   \includegraphics[width=2.5in]{attackmodel.pdf} %\vspace{-0.2cm}
%    \caption{Illustration of on-body backscatter and off-body attack.}
 %   \label{fig:model} %\vspace{-0.3cm}
%\end{figure}

The reminder of this paper is organized as follows. In Section~\ref{sec:motivation}, we first discuss security threats to on-body backscatter and propagation features that motivate our design. Section~\ref{sec:design} elaborates each component in our design framework. Implementation and evaluation are presented in Section~\ref{sec:evaluation}, followed by discussion in Section~\ref{sec:discuss} and literature review in Section~\ref{sec:related}. Finally, Section~\ref{sec:conlude} concludes this work.
\section{Motivation} \label{sec:motivation} %3.3 0.5pp
In this section, we first discuss the potential threats to on-body backscatter communications and argue that an authentication mechanism designed for securing on-body backscatter is critical. Next, we investigate the features of the on-body radio propagation, which motivate our authentication design for on-body backscatter communications.

\subsection{Threat Model}\label{sec:threat} 
Recently, backscatter communications are envisioned to be a promising choice for wearables and implantable devices~\cite{interscatter,onbodyscatter}. On-body backscatter communications bring a bundle of exciting features due to its minimalist. However, the other side of the coin is that the minimalist design of backscatter makes it vulnerable to the presence of malicious attackers. For example, a nearby tag or a powerful active radio emits signals very similar to an on-body tag, and launch user spoofing attacks. Such kind of attacks may lead to serious issues as wearables and implantables are directly related to the human body. Thus, an effective authentication method is necessary. As shown in Fig.~\ref{fig:model}, we consider there has been a backscatter communication link established on the body, in which, the tag transmits data to an on-body receiver by reflecting an on-body link (e.g., from smart watch to smartphone). In this case, a nearby attacker aims to impersonate the tag to send fake data to the receiver. In this work, we consider two types of attackers:~\textit{active attacker} and~\textit{tag attacker}.

\textbf{Active attacker.}An active attacker is an active radio that has all the priori knowledge of the transmitter and the transmitted signal, including the modulation/coding scheme of transmitted packets, the bitrate, code length and the carrier frequency, of on-body tag. With such a powerful knowledge, the attacker can perfectly emulate the backscatter signal to launch spoofing attack. Additionally, the attacker can employ multiple antennas to monitor the channel and simultaneously emit signals using the same technology as the on-body smart devices. In this paper, we consider two types of the active attackers:~\textit{constant power active attacker} and~\textit{powerful active attacker}. 

The constant power active attacker is the radio that can imitate the reflecting path of the on-body backscatter tag by transmitting data at constant power intermittently. As for the powerful active attacker, it is the radio that can transmit varying powers to imitate the dynamic body or on-body antenna movement effects intermittently based on the received signal strength (RSS) variations captured by multiple antennas. Without an authentication mechanism, the receiver demodulates all reflections in the received signals including fake data from the active attacker.

\textbf{Tag attacker.} Tag attacker is an off-body backscatter tag that employs the same backscatter technology as the on-body tag and is capable of reflecting signals to the receiver. To launch spoofing attack, the tag attacker reflects the received signal following the same protocol as the on-body tag while injecting fake data to mislead the receiver.

\begin{figure}[t]
    \center
    \includegraphics[width=3.7in]{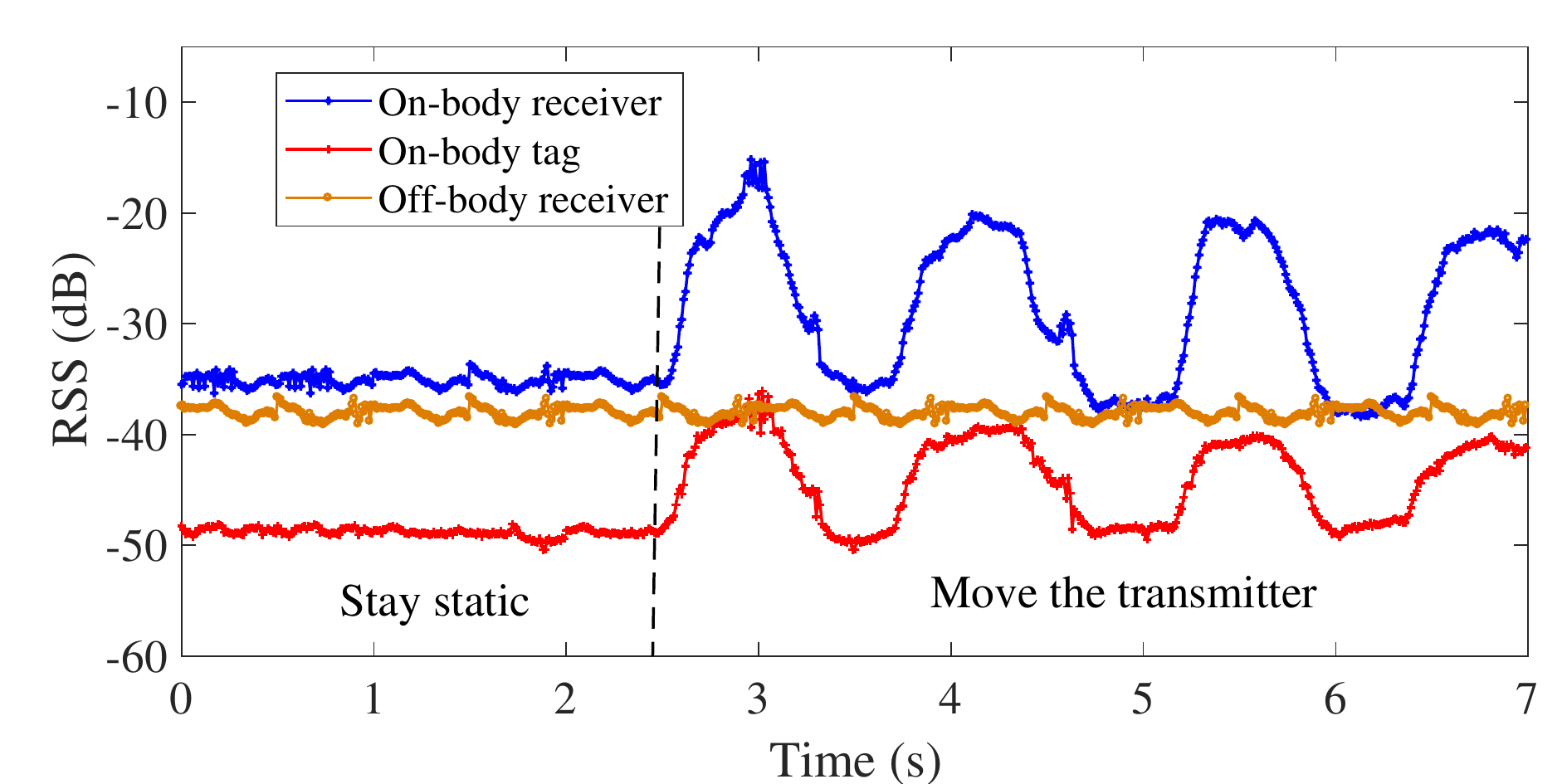} %\vspace{-0.2cm}
    \caption{RSS variations comparison among on-body receiver,  off-body receiver, and on-body backscatter tag.}
    \label{fig:compare} %\vspace{-0.2cm}
\end{figure}

\subsection{On-Body Radio Propagation Signatures}\label{sec:propagation} 
The electromagnetic (EM) waves mainly propagate around the human body surface via diffraction. As a result, the human body, a low loss dielectric medium, usually leads to huge impacts on the radio propagation among on-body devices. According to the creeping-wave and on-body propagation theory~\cite{wait1998ancient,alves2011analytical,wang2017detecting}, if the transmitter and receiver are close to the same body, such as within a few millimeters to centimeters, the wave propagation is dominated by a \textit{creeping} mode. The electric field radiated by a transmitting antenna at a distance $d$ is defined as
\begin{equation}
\\\text{$E$}=\sqrt{{\eta\over{2\pi}}}{\sqrt{P_tG_t}\over{d}}{e^{-jkd}}{W(d,r,{\varepsilon}_r,h_t,h_r)}+\sqrt{{\eta\over{2\pi}}}{\sqrt{P_tG_t}\over{2\pi{r}-d}}{e^{-jk(2\pi{r}-d)}}{W(2\pi{r}-d,r,{\varepsilon}_r,h_t,h_r)},
\end{equation}
where $\eta$ is the vacuum wave impedance, $P_t$ the transmission power, $G_t $ the antenna gain, $k$ the wavenumber in free space, and $r$ the radius of body surface. $W$ is the attenuation function that describes the losses caused by the complex permittivity $\varepsilon_r$ of the human tissue, the curvature $r$ of the body surface and the distances $h_t$, $h_r$ between the body surface and the antennas.
According to Eq.(1), it is obvious that the attenuation of on-body propagation is affected by body surface change and the positions and orientations of the transmitting and receiving antennas, while environmental changes have little impact on on-body propagation.

It has been reported that the path loss is easily effected by the positions of the antennas~\cite{alves2011analytical}. For example, the azimuthal gain is almost 3.5~dB higher for the head than for the waist~\cite{alves2011analytical}. The on-body radio propagation will remain stable if the body stays static. Specifically, based on the measurements~\cite{miniutti2008narrowband,kim2009statistical}, even though the positions of the devices will have significant impact on the antenna gains, the path loss variations of the radio propagation will be less than 4~dB when the human body remains relatively static. On the other hand, even slight body motions would change the body surface curves as well as antenna positions, and thereby cause significant fluctuations in the received signals. 
In contrast, the wave propagation of off-body devices is dominated by direct path and less sensitive to the body surface change and antennas movements but mainly affected by the environment changes.

Based on these results, we consider generating movement-induced variations by intentionally making some movements for the on-body devices. As a result, the on-body links will experience distinct propagation signatures. We can extract these variation features to authenticate the on-body tags. In order to verify the feasibility of our idea, we conduct an experiment by placing a transmitter, a tag, and a receiver on the same body as shown in Fig.~\ref{fig:model}. In our experiments, we use GNURadio/USRP B210 nodes to act as the on-body transmitter , receiver and off-body receiver. All the nodes are powered by 6~V DC input and we use the laptops USB to power these nodes. All the antennas equipped on the on-body transmitter, receiver and off-body receiver are omnidirectional and linear polarized with a gain of 3~dB. We control the USRPs to work at 900MHz. The backscatter tag is implemented according to~\cite{liu2013ambient}. In particular, we use a half-wavelength dipole antenna to reflect the ambient signal. In this case, there are two communication links established on the body, one of which is the backscatter path, and the other one is the direct path from the transmitter to the receiver which is referred to the main path in this paper. Additionally, we also place the off-body receiver 1~m away from the body. First, we keep all these three on-body devices and the body to be static for a period of time and measure the RSS. Then, we keep the tag and receiver to be static and just move the transmitter back and forth for several times. The RSS variations is depicted in Fig.~\ref{fig:compare}. According to the results, we observe when all the devices remain static, all of the received signals stay stable. Whereas, when we move the transmitter, the on-body received signal experiences a large variation, and meanwhile, the reflected signal for the tag varies accordingly. In addition, we also observe that the RSS variations of the tag have the same trend as the main path. On the contrary, the off-body signal still remains relatively stable, which implies that the on-body signals are easily influenced by the on-body transmitter movement, and this signature is not shared by the off-body signals. Based on the above observations, we can exploit these distinct on-body propagation signatures to authenticate whether the received data is from on-body backscatter tags or attackers.

\section{SecureScatter Design} \label{sec:design} %3.3 0.5pp
\begin{figure}[t]
    \center
    \includegraphics[width=4.8in]{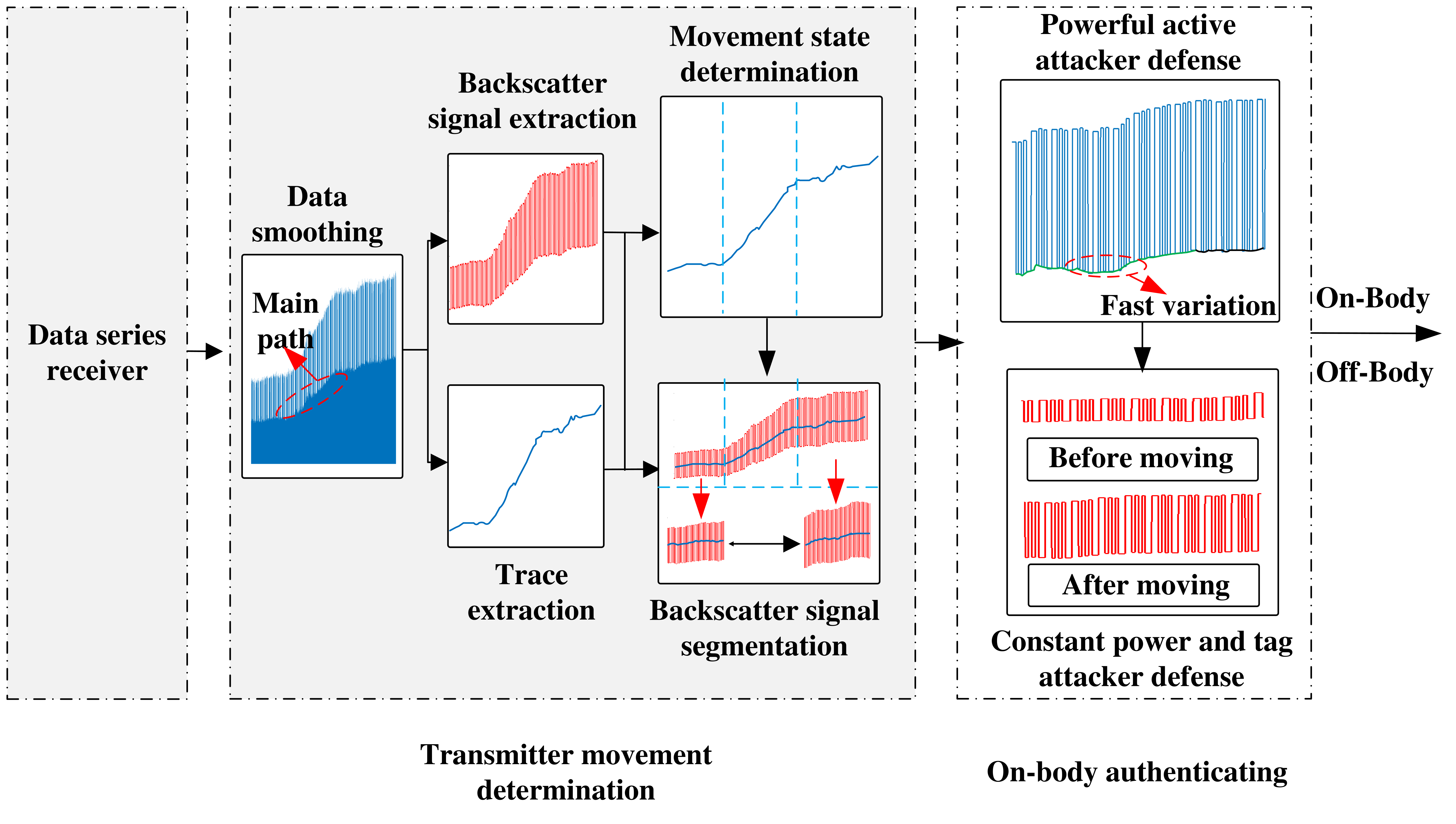} %\vspace{-0.2cm}
    \caption{System flow of SecureScatter.}
    \label{fig:framework}\vspace{-0.3cm}
\end{figure}

The crux of SecureScatter is to take advantage of the received data information and the propagation signatures to authenticate the on-body backscatter tags. Fig.~\ref{fig:framework} illustrates the system flow of SecureScatter. At the beginning, we keep the on-body tag and the receiver to be static. And then preform transmitter movement where we move the transmitter back and forth for several times intentionally. Meanwhile, the receiver logs RSS data and initiates  the authentication process. SecureScatter contains two core components: \textit{transmitter movement determination} and \textit{on-body authenticating}.
\begin{enumerate}
\item \textbf{Transmitter movement determination.} SecureScatter firstly smooths the received signal to remove the fast variations and noise. After that, SecureScatter extracts the backscatter signal from the received data, and then fits the main path signal, i.e., the trace of the signal from the transmitter to the receiver. According to the trace, we can obtain the power changing states of the main path received signal. Based on the trace, SecureScatter can identify the variations of the received signal and then segments the extracted backscatter signal into different groups that correspond to different movement states.
\item \textbf{On-body authenticating.} Based on the backscatter signal segments obtained by the on-body transmitter movements determination, SecureScatter firstly removes effects caused by the powerful active attackers and body movements. After that, the receiver authenticates the on-body tag by comparing the power variations in segments to defend constant power active attackers and tag attackers in each group.

\end{enumerate}

%\begin{figure}
%    \begin{minipage}[t]{0.2\textwidth}
%        \centering
%        \includegraphics[width=1\textwidth]{realuse1}
%        
%    \end{minipage}
%    \hspace{20ex}
%    \begin{minipage}[t]{0.2\textwidth}
%        \centering
%        \includegraphics[width=1\textwidth]{realuse2}
%    \end{minipage}
%    \caption{The typical user scenario of our system. The left-hand figure shows the device worn on the chest. The right-hand figure shows the approximate position on the rib cage and near the diaphragm. In practical usage, the device should be worn inside the clothes to capture a high quality signal. It is reasonable because most of the chest worn wearables on the market need to contact the skin.}
%    \label{realuse} 
%\end{figure}

\subsection{Transmitter Movement Determination}\label{sec: gesture deyermination} 
The first step of SecureScatter is to determine the variation components caused by on-body transmitter movement. To this end, SecureScatter needs to preprocess the received signal to remove noise first. After obtaining the smoothed data, SecureScatter extracts the trace of the main path to determine movement states and then divides the backscatter signal into different segment groups. 
\begin{figure}[t]
	\centering 
	\begin{minipage}[t]{0.2\textwidth}
		\centering
		\subfigure[Received data.]
		{\includegraphics[width=2.2in]{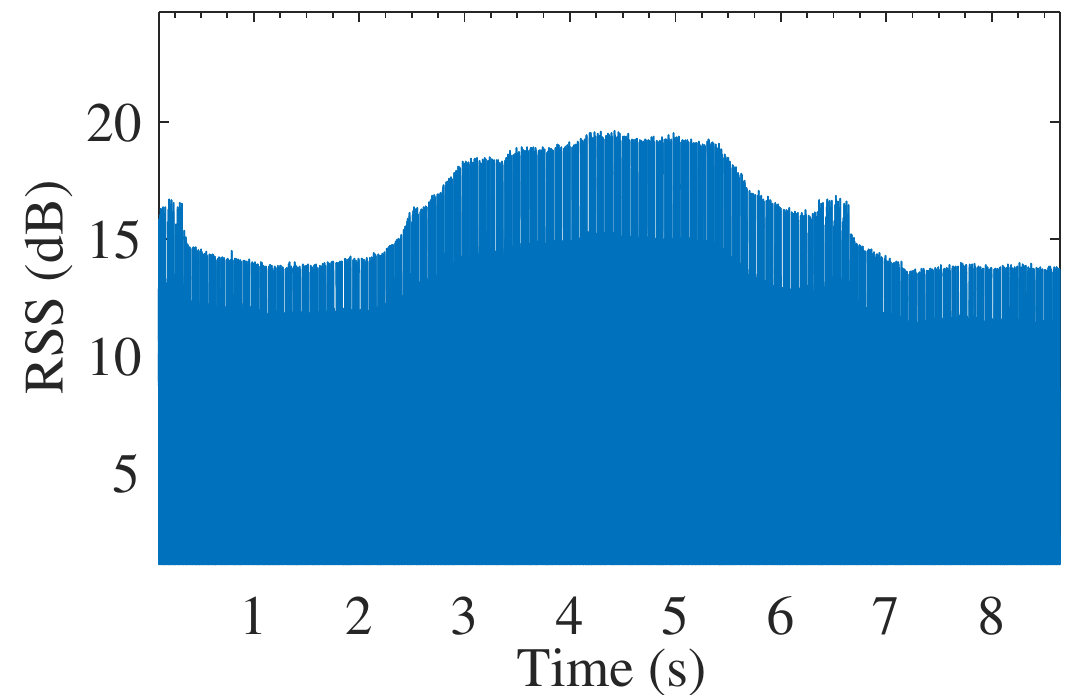}}
	\end{minipage}
	\hspace{20ex} 
	\begin{minipage}[t]{0.2\textwidth}
		\centering
		\subfigure[Noise and fast-varying components.]
		{\includegraphics[width=2.2in]{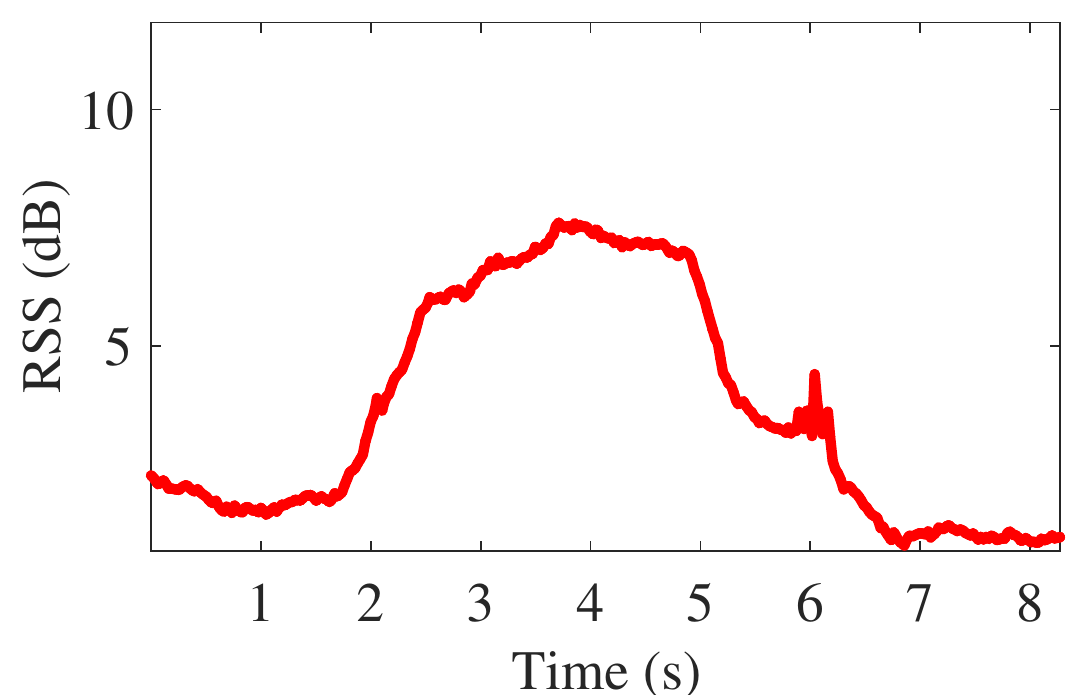}}
	\end{minipage}
	\hspace{20ex}  
	
	\begin{minipage}[t]{0.2\textwidth}
		\centering
		\subfigure[Backscatter signal extraction.]
		{\includegraphics[width=2.2in]{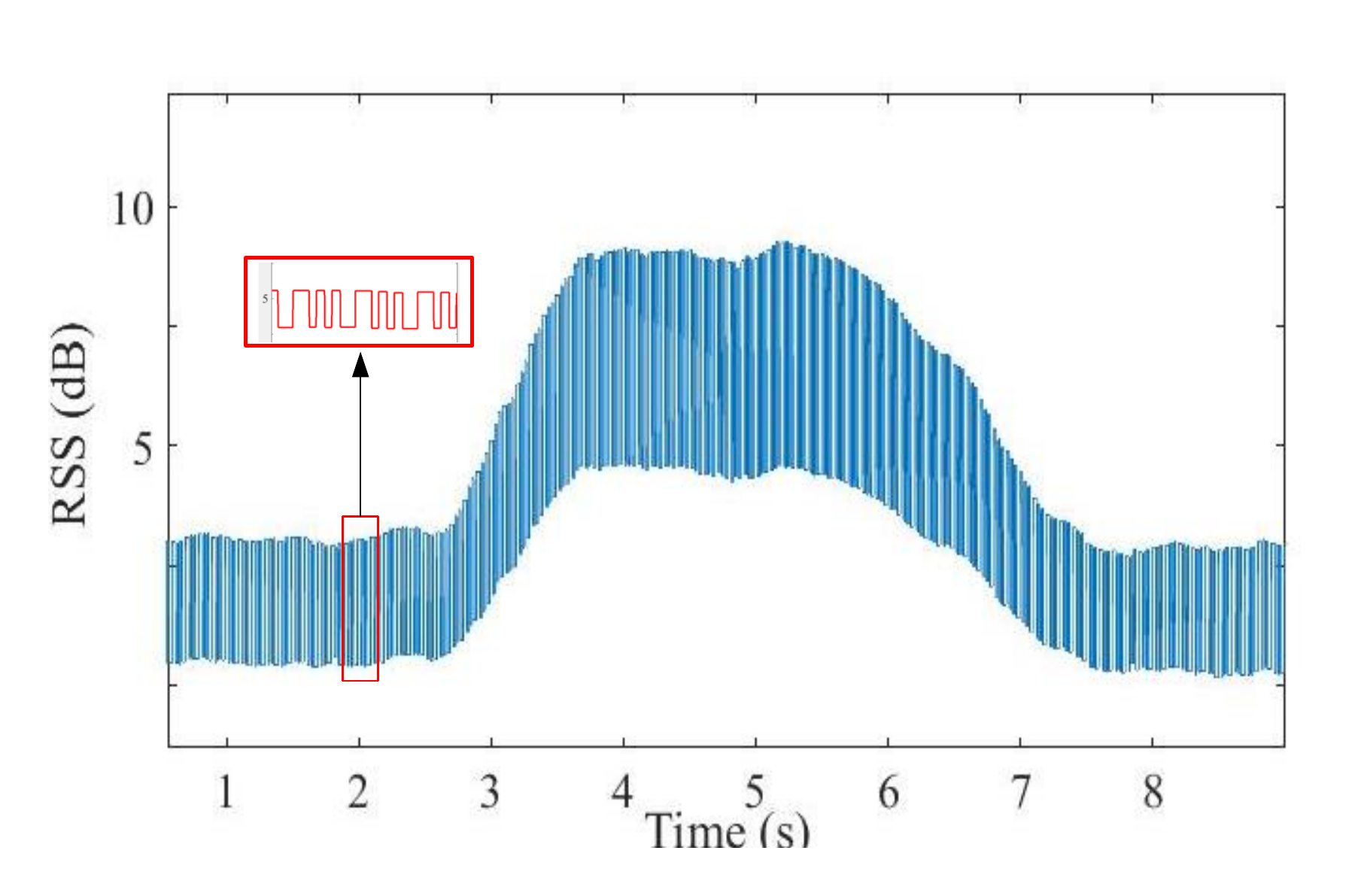}}
	\end{minipage}
	\hspace{20ex}    
	\begin{minipage}[t]{0.2\textwidth}
		\centering
		\subfigure[The trace of the main path extraction.]
		{\includegraphics[width=2.2in]{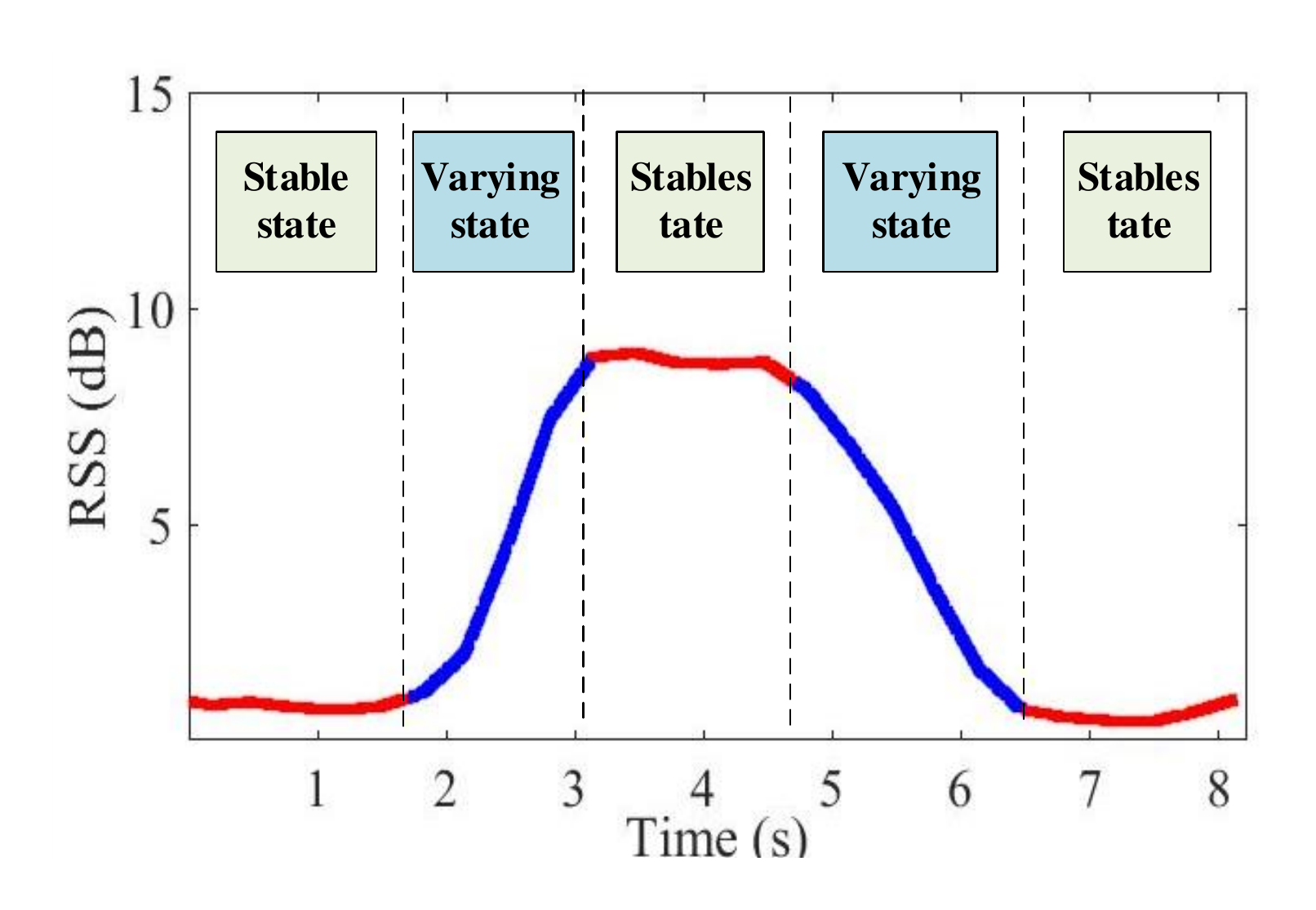}}
	\end{minipage}
	\hspace{20ex}       
	\caption{Received data process.}
	\label{fig:original}\vspace{-0.3cm}
\end{figure}
\textbf{Data smoothing.} Based on the principle of the backscatter, we can learn that the received signals consist of multiple components, including the reflection signal from the tag or the attacker, the noise caused by tag circuit and environment, as well as the fast-varying signal generated by the transmitter~\cite{liu2013ambient} as shown in Fig.~\ref{fig:original} (b). To extract the backscatter information and main path signal, SecureScatter needs to filter out variations that are irrelevant to them. To this end, SecureScatter first employs a filter to screen out the low-frequency components caused by tag circuit and environment.  After that, we employ a moving window of 50 samples to eliminate the fast-varying signal and obtain the smoothed data.
%\begin{equation}
%\\x(n)=\\{1\over \\N} \sum_{i=n}^{n+N} |{x(i)}|
%\end{equation}

\textbf{Backscatter signal and trace extraction.} Without loss of generality, we tailor the detailed algorithm for the widely-adopted backscatter technology~\cite{liu2013ambient,huang2018toward,widescatter,nicscatter}, in which the backscatter signal reflects the main path signal intermittently by acting as an additional reflection path. Thus, the receiver can treat the backscatter signal as amplitude modulation (AM) and demodulate it by detecting the variations of the power.  SecureScatter employs a low-pass filter to demodulate the smoothed data and then separates the backscatter signal from the received data as shown in the Fig.~\ref{fig:original} (c).

Based on the discussion in Section~\ref{sec:motivation}, we observe that the trace of the main path indicates the average RSS from the transmitter to the receiver. Thus, we need to extract the trace to determine the state of the transmitter first. In order to extract the trace of the main path, SecureScatter has to recover the power of the main path. On the basis of the smoothed data samples, we observe that the backscatter signal have an obvious boundary with the main path. Accordingly, SecureScatter removes the power of the extracted backscatter signal first. Thereby, the received data only contains the main path signal after this operation. Then, SecureScatter calculates the RSS variations of these left signal adopting a moving window, in which we consider the average signal strength of each window to indicate the trace. Thus, SecureScatter can obtain a raw trace for the main path signal.

\begin{figure}[t]
    \center
    \includegraphics[width=4.5in]{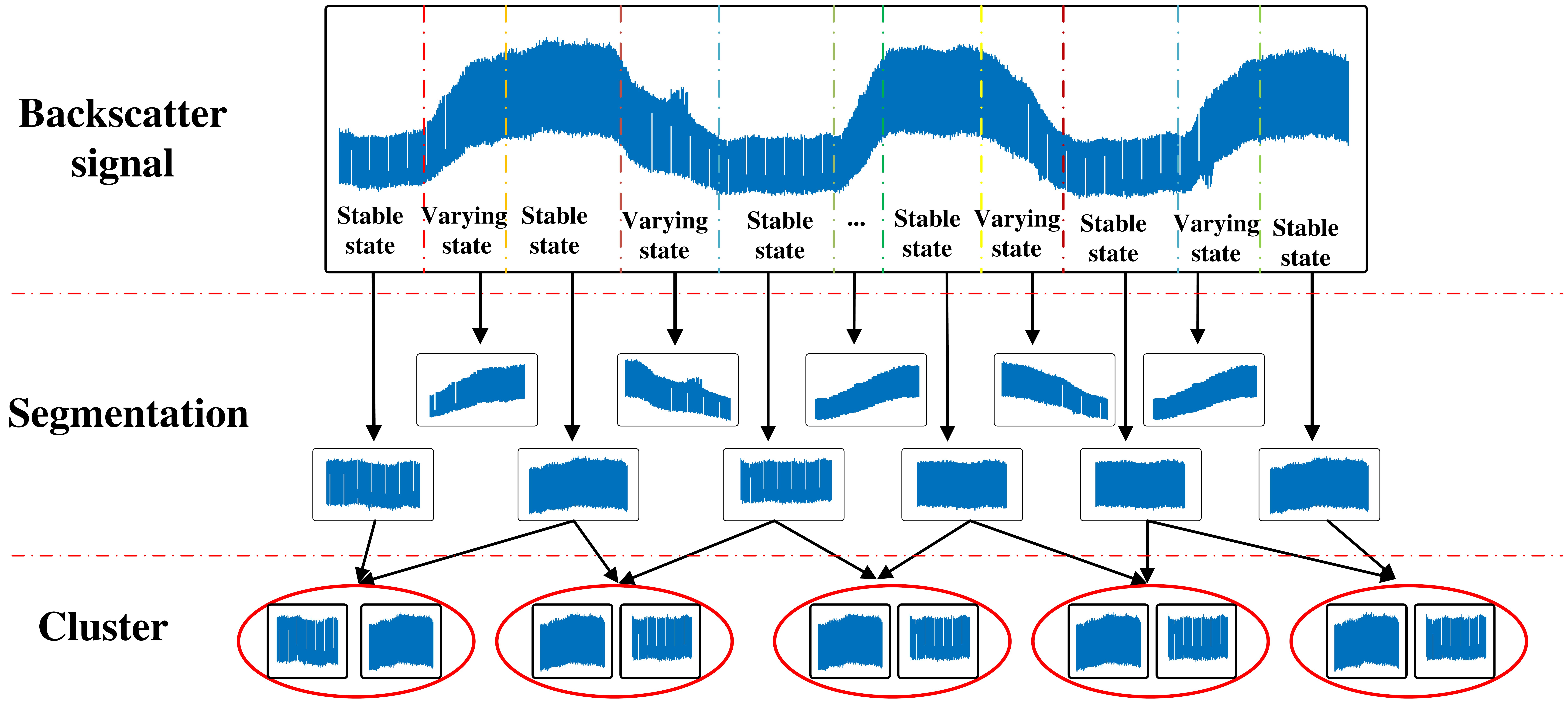} %\vspace{-0.2cm}
    \caption{Signal segmentation and clustering. Based on the trace of the trace of the main path, the backscatter signal can be divided into stable states and varying state. Then the two adjacent stable states are clustered into the same group. }
    \label{fig:cluster}%\vspace{-0.5cm}
\end{figure}
\textbf{Movement state determination.} Recall that the main path signal occupies the main ingredients in the received sample series, and the variations of the tace of the main path mainly reflect the changes of the states when moving the transmitter. When the main path trace stays stable, the devices on the body are more likely to stay in static. Conversely, if the trace of the main path has a large variation, it implies that the transmitter is moving. Hence, SecureScatter can perform movement state determination based on the trace of the main path.

Based on the above signatures, SecureScatter exploits a slope-based mechanism to detect the movement states. SecureScatter first smooths the trace of the main path again so as to eliminate the fluctuation caused by the dynamic on-body channel, and then computes the slope of the trace and normalize it as follows,
\begin{equation}
\\\text{slop}(n)=\text{Normal}\left|{\sum_{i=n+N}^{n+2N} |{x(i)}|-\sum_{i=n}^{n+N} |{x(i)}|\over N^2}\right|,
\end{equation}
where $N$ represents the smooth interval, which is set to $N={w\times sample rate \over bitrate}$. $w $ is a  constant coefficient and we set $w=1.2$ in our experiments. After deriving the slopes series, SecureScatter sets a threshold to define the movements states for the trace. Then, SecureScatter retrieves the slope series. In order to lower the contingency, if finding that the value of slope is smaller than the threshold and existing over three intervals that less than the threshold in front of this value simultaneously, SecureScatter treats these intervals as a stable state and gives a mark on it. On the basis of this process, SecureScatter repeatedly retrieves all the slope series and records all the marks. We set the threshold as 10~dB in our experiments empirically.

Performing slope-based mechanism can only yield a raw movement state determination as the dynamic body and environment can also lead to large slope variations. Thus, we have to further remove these effects by state selection. Recall that when the body remains static, the variations are less than 4~dB~\cite{miniutti2008narrowband,kim2009statistical}. However, the movement of the transmitter will lead to large variations~\cite{alves2011analytical}. Besides, we also find that each movement duration always locates at the center of the two adjacent stable states. Thus, SecureScatter compares the average RSS of these two adjacent stable states and if the difference of these two stable states is larger than 4~dB~\cite{miniutti2008narrowband,kim2009statistical}, these three states are deemed to be caused by transmitter movements otherwise considered as caused by dynamic effects and are discarded. In this way, SecureScatter removes the dynamic channel effect and select the reliable states based on the variation of the trace powers as shown in Fig.~\ref{fig:original} (d).
\begin{figure}[t]
    \center
    \includegraphics[width=3.1in]{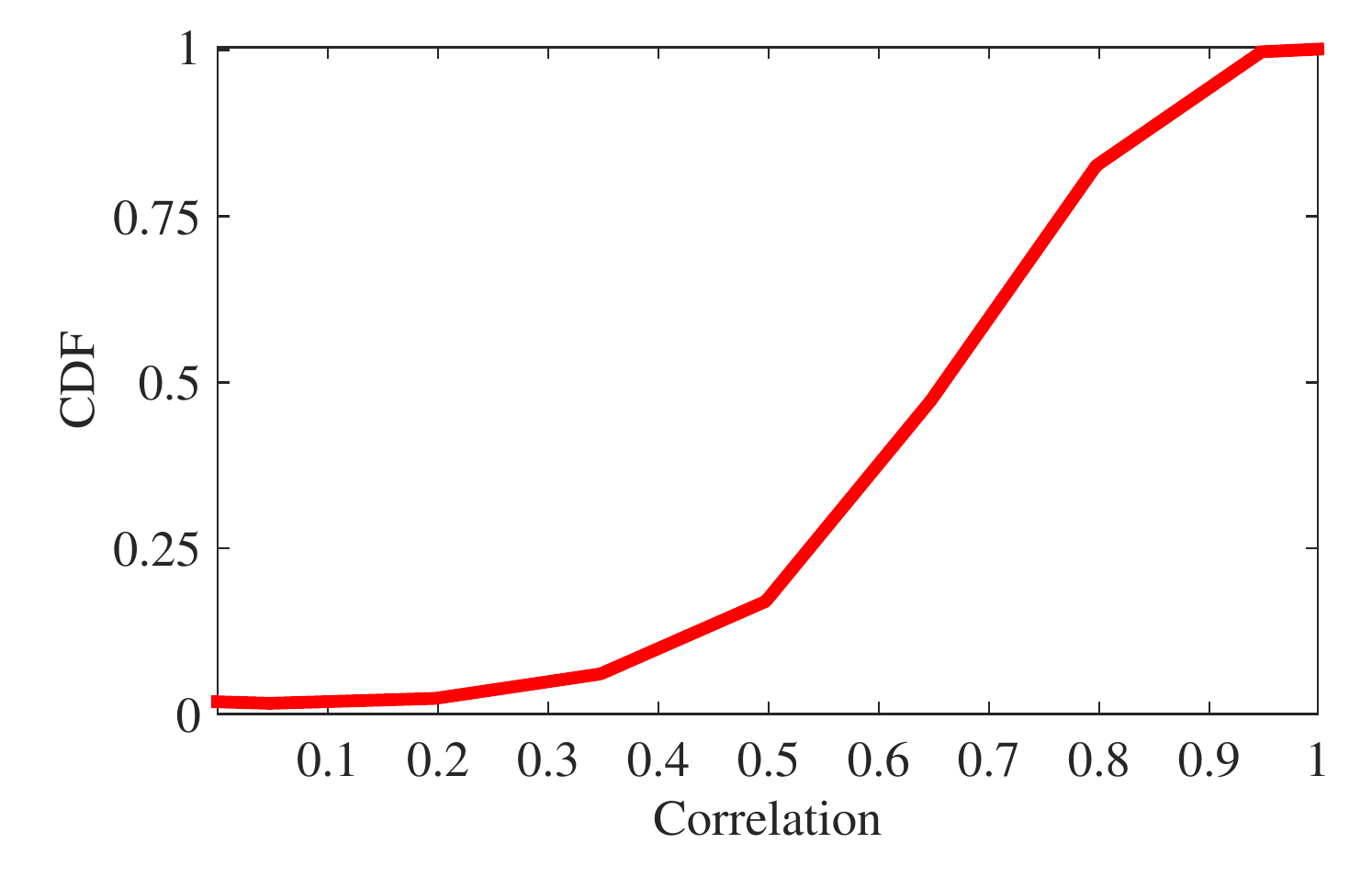} %\vspace{-0.2cm}
    \caption{CDF of correlation between backscatter signal and main path.}
    \label{fig:correlation} \vspace{-0.3cm}
\end{figure}

\textbf{Backscatter signal segmentation.} By far, SecurScatter has obtained the movement states of the trace of the main path. Then, we need to determine the states of the backscatter signal. According to the priori knowledge that the trace of the main path indicates the states of the on-body transmitter, we can detect the states of the backscatter signal and segment them into different groups based on the trace of the main path. 

In order to segment and cluster the backscatter signal, the first step is to match the trace of the main path with the extracted backscatter signals. After that, we segment the backscatter signals based on the trace states of the main path obtained by the slope-based mechanism and the states selection. As a result, we can obtain a large number of segments of backscatter signals states in which the movements states locate at the center of the stable states. In the next step, we cluster the stable states into different groups. Based on the trace of the main path, we can see that the two stable states of the backscatter signal in each group appear before or after the transmitter movements. The clustering results can be seen in Fig.~\ref{fig:cluster}.

\subsection{On-Body Authenticating}\label{sec:detecting} 
By far, SecureScatter has selected all the segments in the groups, in each of which the two segments represent the stable states before and after moving on-body the transmitter. In the following step, SecureScatter needs to first perform authentication to remove the powerful active attacker who can transmit varying powers first, and then defend against the constant power and tag attacker.

\textbf{Powerful active attacker defense.} The powerful active attackers who can transmit different levels of powers to mislead the receiver based on detecting the RSS variations with multiple antennas. The RSS variations correlation of the backscatter signal and main path with multiple data series, as shown in Fig.~\ref{fig:correlation} can verify that they are highly correlated with each other. It implies that if the transmitter moves, the RSS variations of backscatter signal will change with the main path accordingly. Conversely, the off-body receivers are not sensitive to the RSS variations as the on-body receivers when moving the on-body transmitter. Thus, the powerful attackers who transmit varying powers based on received RSS variations cannot always follow the transmitter movement precisely and timely, which will lead fast-varying to each stable states. 
\begin{figure}
	\begin{minipage}[t]{0.48\textwidth}
		\centering
		\includegraphics[width=1\textwidth]{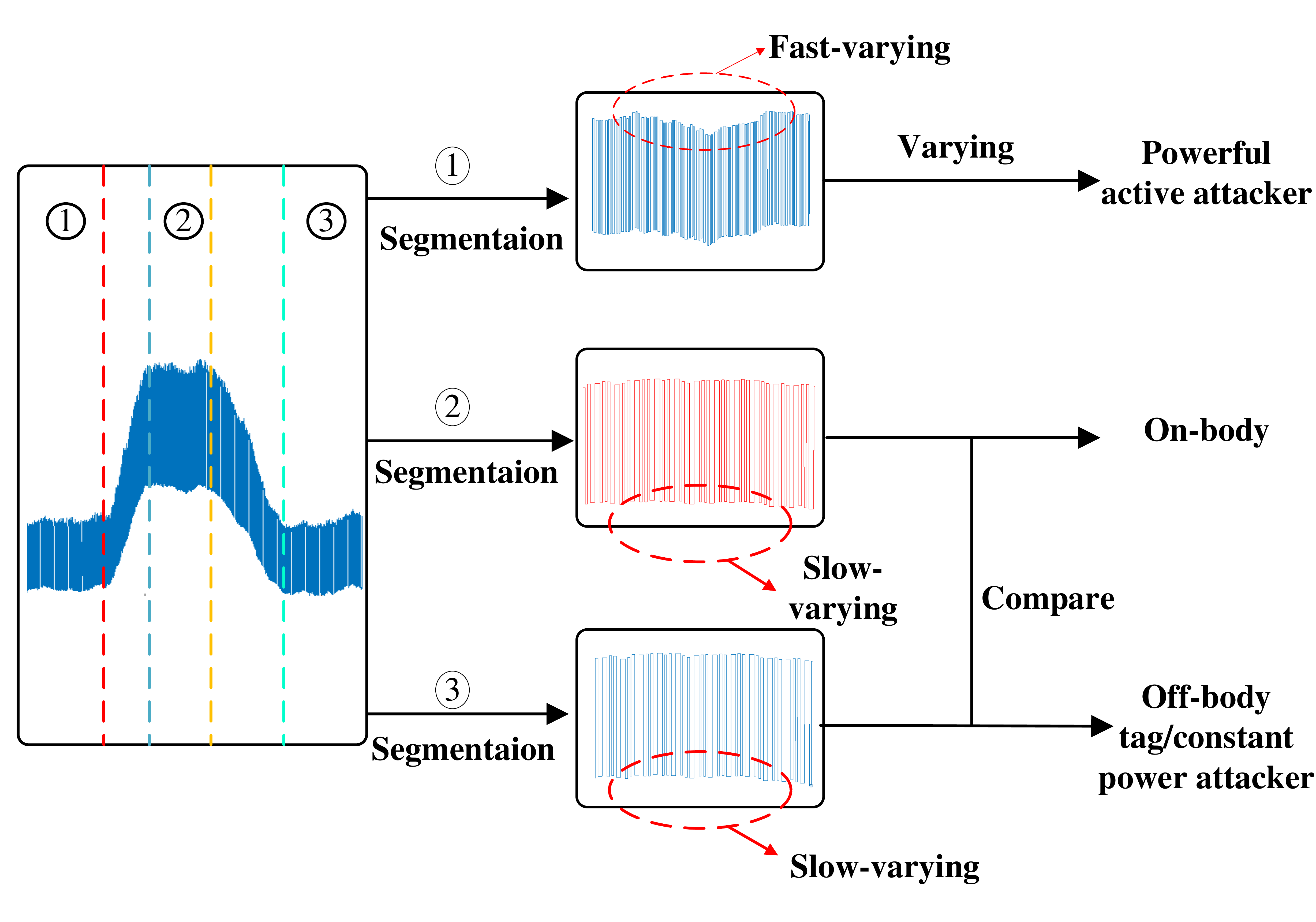}
		\caption{On-body authenticating.} 
		\label{fig:varying}
	\end{minipage}
	\hspace{1.7ex}
	\begin{minipage}[t]{0.48\textwidth}
		\centering
		\includegraphics[width=0.9\textwidth]{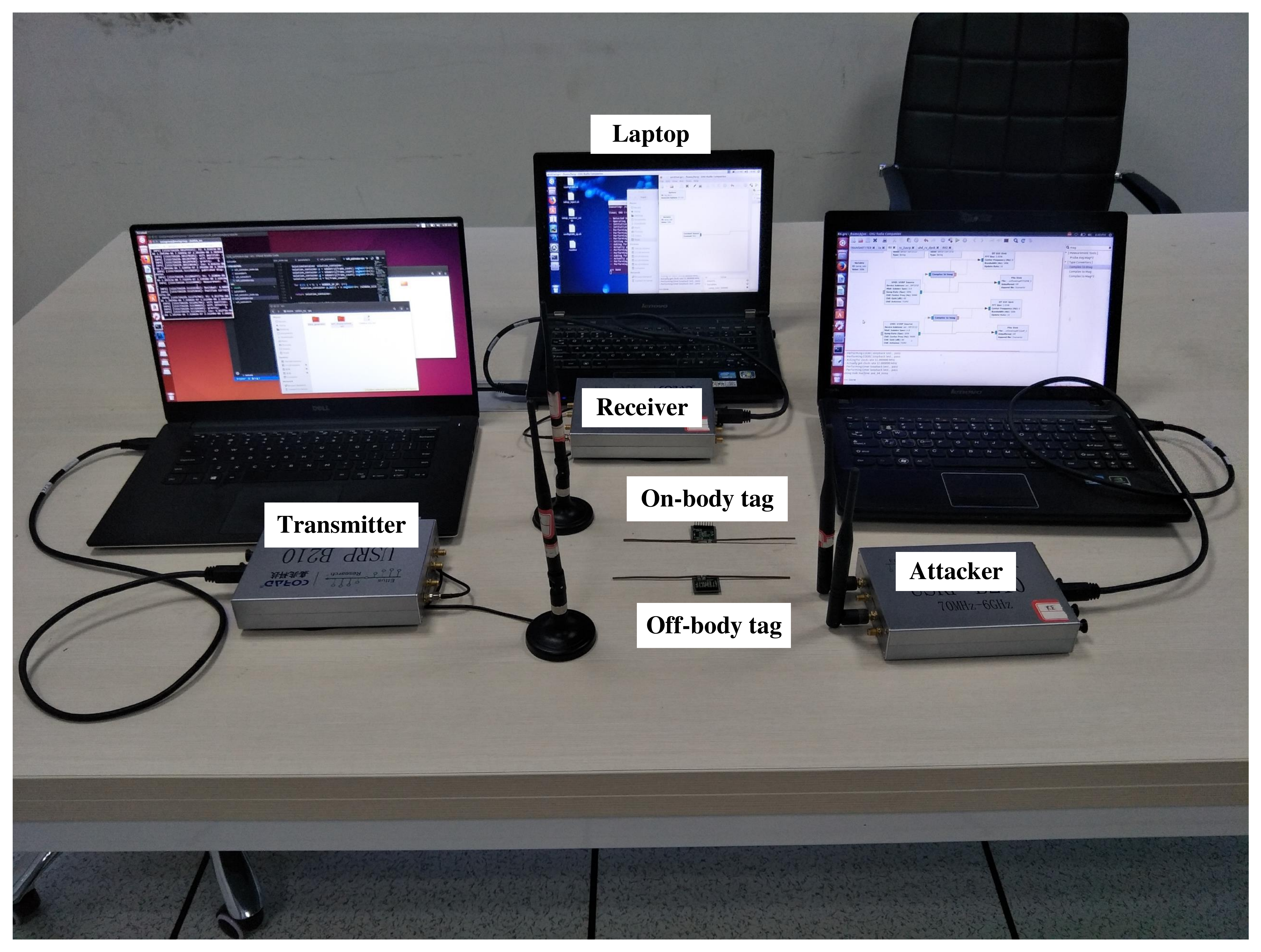}
		\caption{Hardware.}
		\label{fig:hardware} 
	\end{minipage}
\end{figure}
The slight movements of the body, such as hand movements, head shakes or even breath are inevitable and will lead different variations to the backscatter signal. However, based on the experiments shown in Figure~\ref{fig:compare}, we observe when the on-body tag and receiver stay relatively static, and just slightly move the body, such as head shakes, or breath, after performing data smoothing to remove the noise and dynamic body effects, the RSS variations will be less than 3~dB. The result of this experiments matches the measurements in~\cite{kim2009statistical}. It implies if the variation in each segment is larger than 3~dB, we can determine that the variation in this segment is caused by a powerful active attacker. Thus, in order to detect the variation and defend the powerful active attacker, SecureScatter exploits the variance of each segment by,
\begin{equation}
\\Var=\text{Normal}\left|{\sum_{i=1}^{N}( |{x(i)}|-{{1\over N}\sum_{i=1}^{N} |{x(i)}|})^2\over N}\right|,
\end{equation}
where $Var$ represents the variance of each stable state segment in the group as shown in Fig.~\ref{fig:cluster}, and $N$ represents the length of this segment. Based on the variance, we can obtain the change of each segment. Then, we set a threshold to discriminate whether this segment is fast- and slow-varying signal shown in Fig.~\ref{fig:varying}. If the variance is larger than the threshold, SecureScatter deems that the signal in the segments is a fast-varying signal and from a powerful attacker, otherwise it is a slow-varying signal. Thereby, SecureScatter removes the effects of the powerful attacker, and then if it is a slow-varying signal, we consider the tag may be attacked by the constant power attacker or the tag attacker and need to perform a further on-body detection procedure. Based on the experiments in Fig.~\ref{fig:compare} and the measurement result in~\cite{miniutti2008narrowband,kim2009statistical}, we set the threshold as 2.5~dB in our experiments.
\begin{figure}
	\begin{minipage}[t]{0.415\textwidth}
		\centering
		\includegraphics[width=1\textwidth]{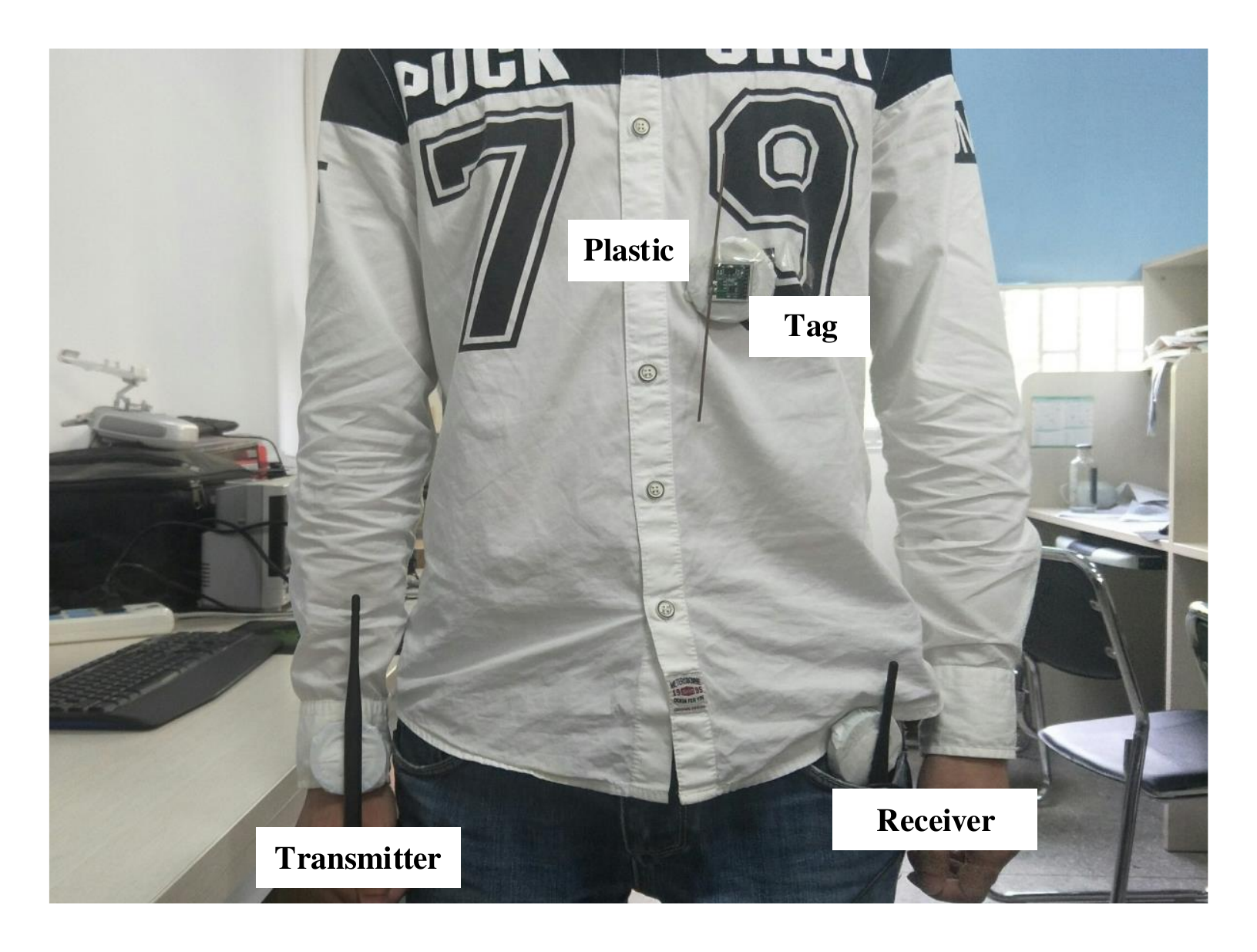}
		\caption{On-body device placement.} 
		\label{fig:scenario}
	\end{minipage}
	\hspace{1.7ex}
	\begin{minipage}[t]{0.48\textwidth}
		\centering
		\includegraphics[width=0.9\textwidth]{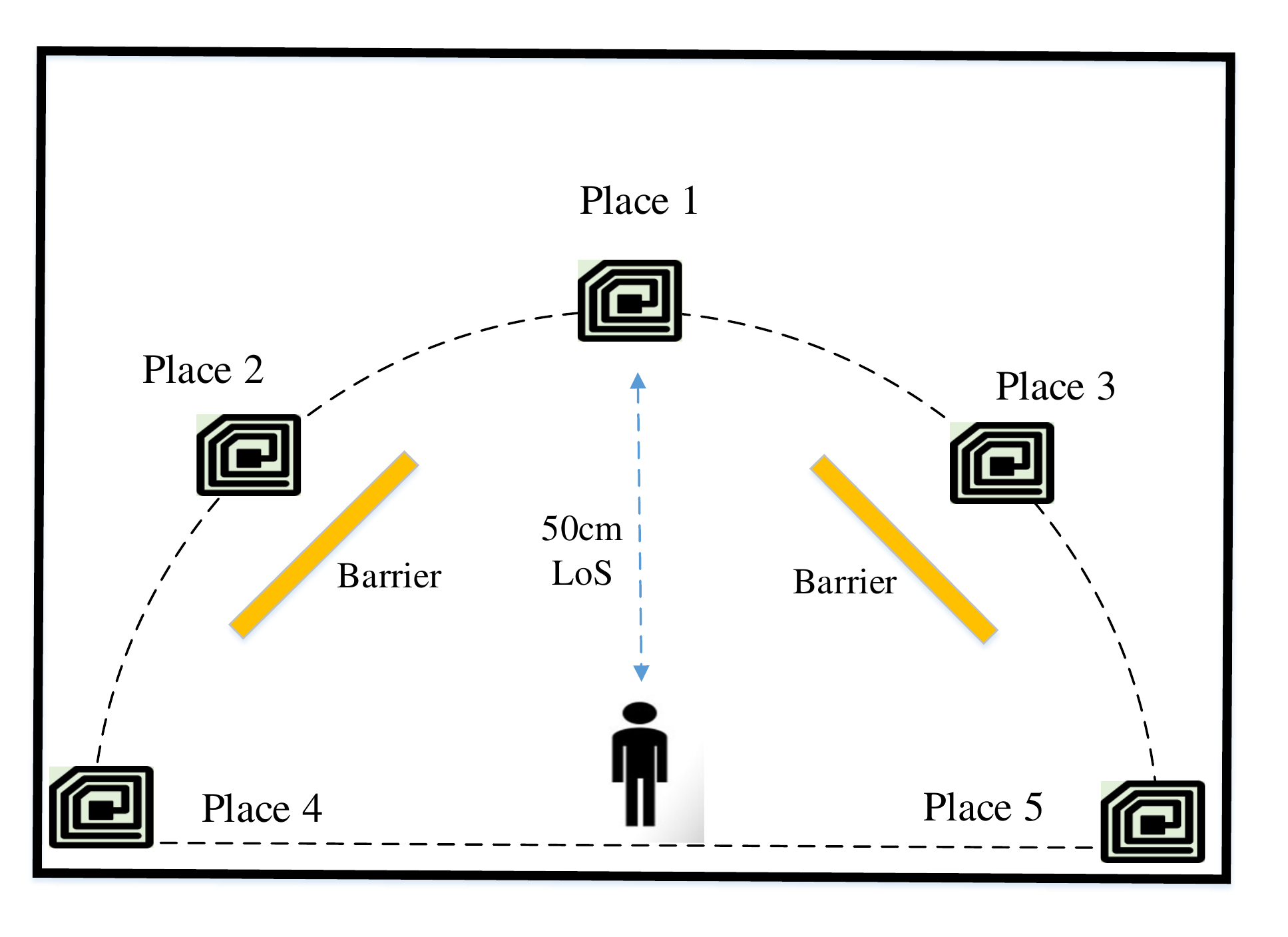}
		\caption{Different places of the attacker.}
		\label{fig:place} 
	\end{minipage}
\end{figure}

\textbf{Constant power and tag attacker defense.} The above process has removed the powerful active attackers with fast variations. The next step for SecureScatter is to defend the constant power attackers and off-body tag attackers. According to the priori knowledge that when moving the on-body transmitter, the RSS of backscatter signal experiences a large variation before and after this movement. On the contrary, the constant power attackers and off-body tags lack these features because they are insensitive to the on-body transmitter. Thus, in order to identify the on-body tag, SecureScatter compares the changes of the two stable states segments in each group. To this end, SecureScatter first computes the average power strength. If the difference between these two segments is larger than the threshold, SecureScatter determines that the backscatter signals are from the tag on the body. Otherwise, the backscatter signal is considered as from an attacker. In order to improve the reliability of the results, SecureScatter can authenticate the tags with multiple groups. Based on our experiments and~\cite{miniutti2008narrowband,kim2009statistical}, the variations must be larger than 4~dB and thereby we set the threshold to be 4.5~dB.

%\begin{figure}[t]
%\center
% \includegraphics[width=3.3in]{figs/framework_1}\vspace{-0.2cm}
% \caption{RSS variance comparison between on- and off-body propagations.}
%\label{fig:framework_1}\vspace{-0.3cm}
%\end{figure}
%\input{implementation}
\section{Implementation}\label{sec:implementation}

In this section, we describe the implementation of SecureScatter in detail as follows.

As illustrated in Fig.~\ref{fig:hardware}, the prototype of SecureScatter is implemented using two backscatter tags and three GNURadio/USRP B210 nodes. All the USRP B210 nodes are powered by separate laptop input~\cite{B210}. We use laptops' USB 3.0 ports to power up these nodes and all the laptops are also battery-powered. The backscatter tags are implemented according to~\cite{liu2013ambient}. In particular, we use half-wavelength dipole antennas to reflect the ambient signal. The half-wavelength dipole antenna is made of copper wire with a diameter of 0.9~mm. We tailor the antenna design to allow tags working at both 900~MHz and 2.4~GHz. One tag is put on different places of the body as the on-body tag, and the other tag is placed away from the body as the off-body tag attacker. In order to emulate the off-body tag attacker, both tags have the same coding scheme, code length and the same bitrate in our experiments. One USRP node equipped with two antennas acts as an active attacker, who monitors RSS variations with one of the antennas while transmitting fake data using the other antenna. The other two USRP nodes are used as the on-body transmitter and receiver, respectively, whose antennas are placed on the body of one user. It is worth noting that all the antennas equipped on the on-body transmitter and receiver are omnidirectional and linear polarized with a gain of 3~dB.

Fig.~\ref{fig:scenario} shows the on-body device placement of SecureScatter. We maintain the distances between the antennas of on-body devices and body to be smaller than 2~cm, which fits the scenario of most wearables. In particular, we place the antenna of the transmitter at the wrist to act as a smartwatch or wristband and the antenna of the receiver in a pocket as a smartphone, respectively. In our experiments, we use plastic blocks with a height of 1.5 cm as underlay of antennas to keep them from directly in contact with the skin. The on-body tag is placed at different places of the body. Besides, the active attacker can be deployed at different directions and distances from the legitimate user.

\section{Evaluation}\label{sec:evaluation} 

We have conducted a series of experiments to evaluate the performance of SecureScatter in both static and dynamic environments. In the static environment, volunteer users are asked to stay stationary in the lab during the test. Then we evaluate the experiments when the attacker is at different distances and directions as illustrated in Fig.~\ref{fig:place}. In the dynamic environment, users are asked to do some slight motions during the experiments so as to imitate real body movements. Furthermore, two nearby users perform their daily activities by walking around at different distances. We evaluate SecureScatter in different time with different users. We also evaluate SecureScatter with different backscatter tags that work at 900~MHz and 2.4~GHz bands that are commonly used by wearables and IoTs.

\textbf{Metrics}. We employ the following metrics to evaluate the performance of our system.
\begin{itemize}
    \item \textbf{TP rate.} TP rate is defined to be the ratio of the number of segments in which the on-body backscatter tag  is correctly authenticated to the total number of segments.
    \item \textbf{FP rate.} FP rate is the ratio of the number of segments in which the off-body attacker is falsely recognized as being on-body to the total number of segments.
\end{itemize}
\subsection{Static Environment}\label{sec:static_experiments} 
First, we show how SecureScatter performs in the office environment where users keep stationary. We evaluate SecureScatter with different distances from attackers and further evaluate the cases in which the attacker is placed in different directions.

\begin{figure}[t]    %    \center 
    \subfigure[FP rates]
    {\includegraphics[width=2.7in]{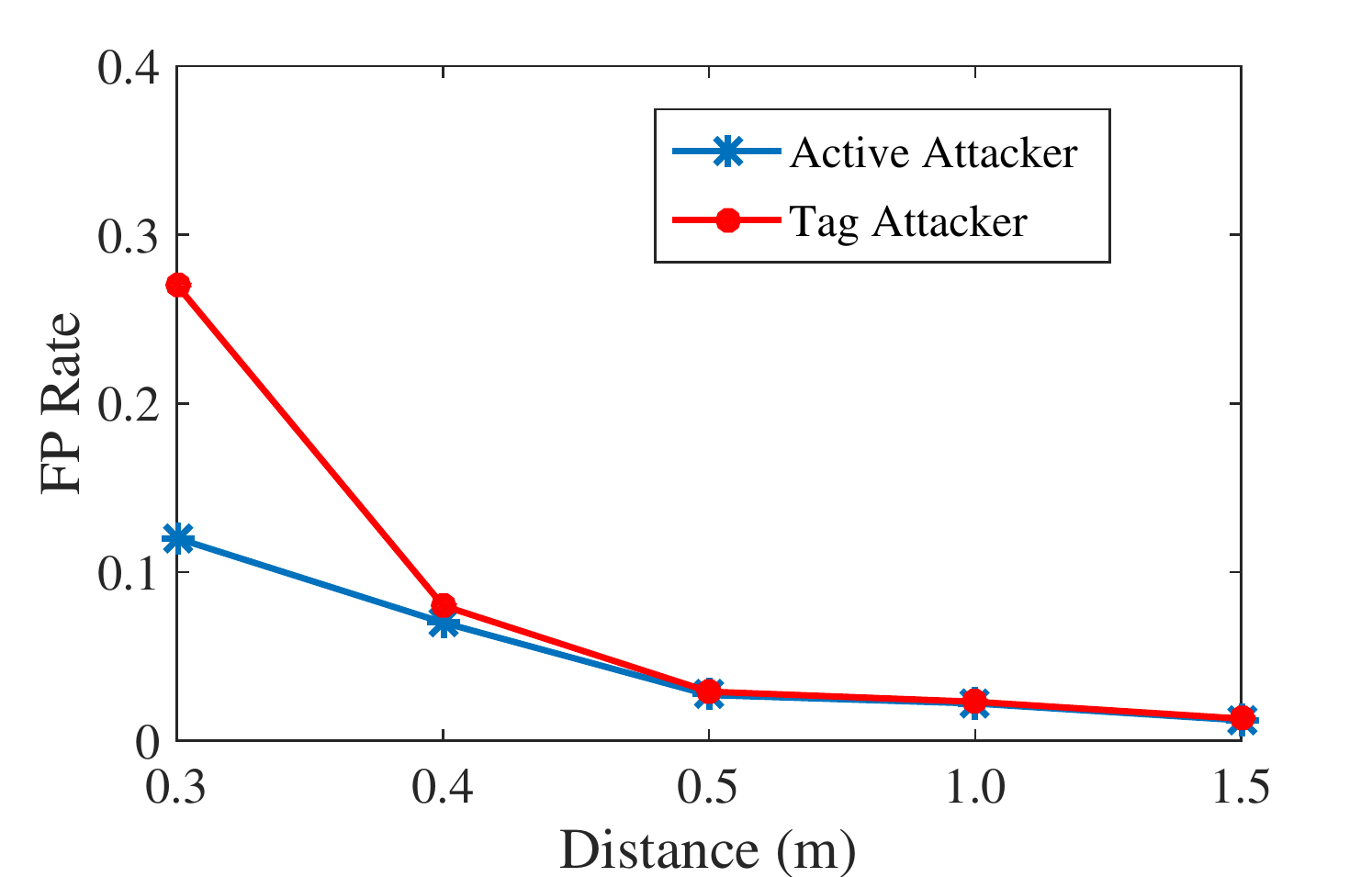}}
    \subfigure[TP rates.]
    {\includegraphics[width=2.7in]{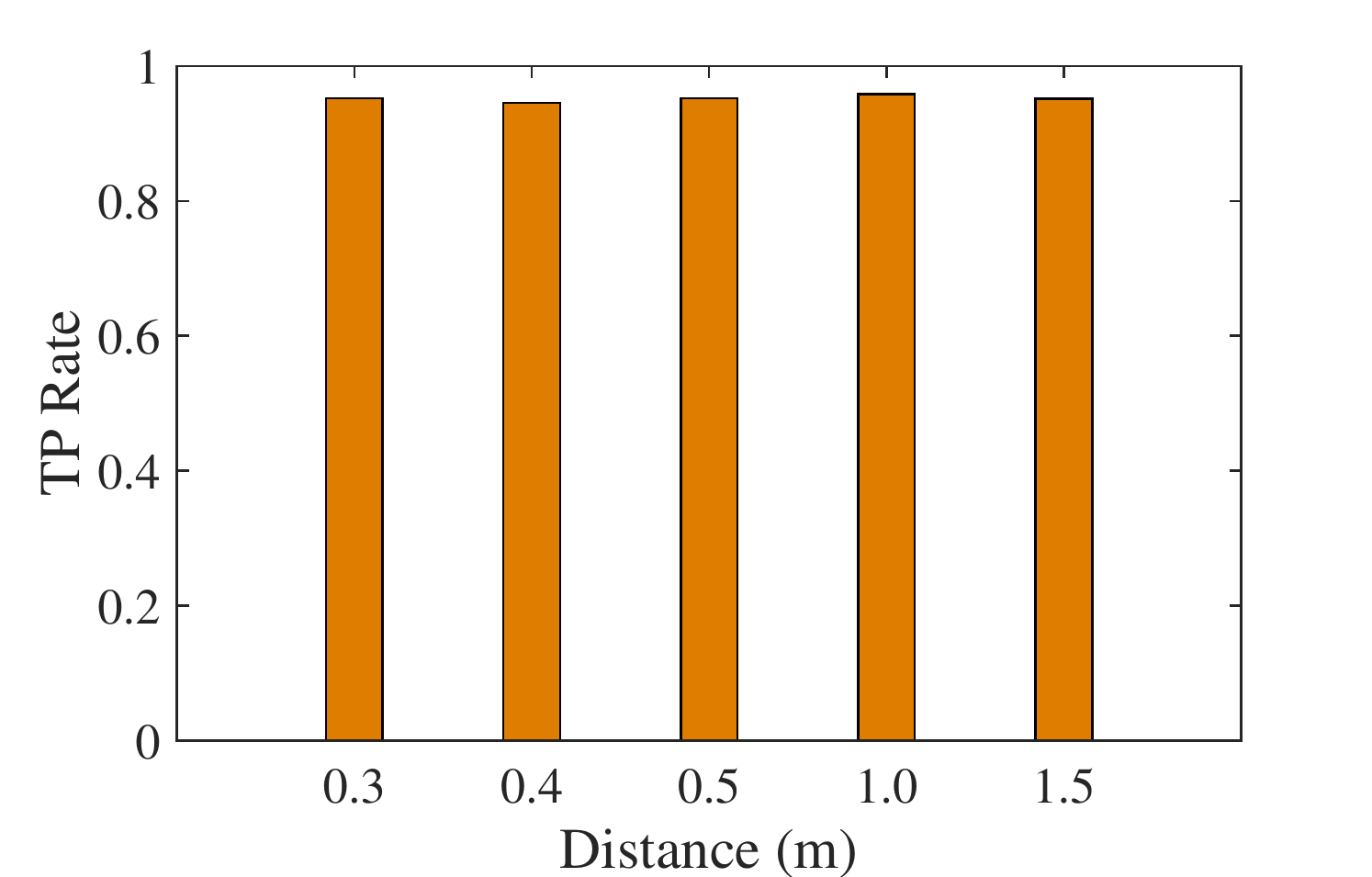}}
    \caption{TP and FP rates at different distances in static environment.}
    \label{fig:static_distance}\vspace{-0.3cm}
\end{figure}
\textbf{Scenarios}. 
In the first set of the experiments, we first evaluate the performance of SecureScatter by varying distances between the active attacker and the user in static environments.  To this end, we place the tag on the chest, and then the antenna of the transmitter is placed in hand, while the antenna of the receiver is placed in the pocket position. When testing, we slowly move the antenna of the transmitter back and forth for several periods, meanwhile, the other parts of the body are kept relatively static. Besides, for the active attacker, a constant power active attacker who can transmit constant power intermittently and a tag attacker are placed at the same direction but different distances away from the user.

In the second set of the experiments, we follow the same settings for on-body devices as the first set of the experiments while the only departure is that we change the directions of the attacker. In particular, we place the attacker in five different positions 0.5~m away from the user as shown in Fig.~\ref{fig:place}, where the place~1 is in line-of-sight LoS, and place~4 and~5 are the direction aside the user. The place~2 and~3 are sheltered by the wood barriers, so that the effects of the  barriers on this system can be taken into account. 

In the third set of the experiments, in order to enhance the reliability of the system, the user is asked to increase the number of the times of the movements instead of only one time as the previous experiments. Then we combine the results in all the number of the times for the movement to authenticate tags. SecureScatter decides the authentication results by majority vote based on the number of the times.

\begin{figure}[t]
	\subfigure[FP rates.]
	{\includegraphics[width=2.7in]{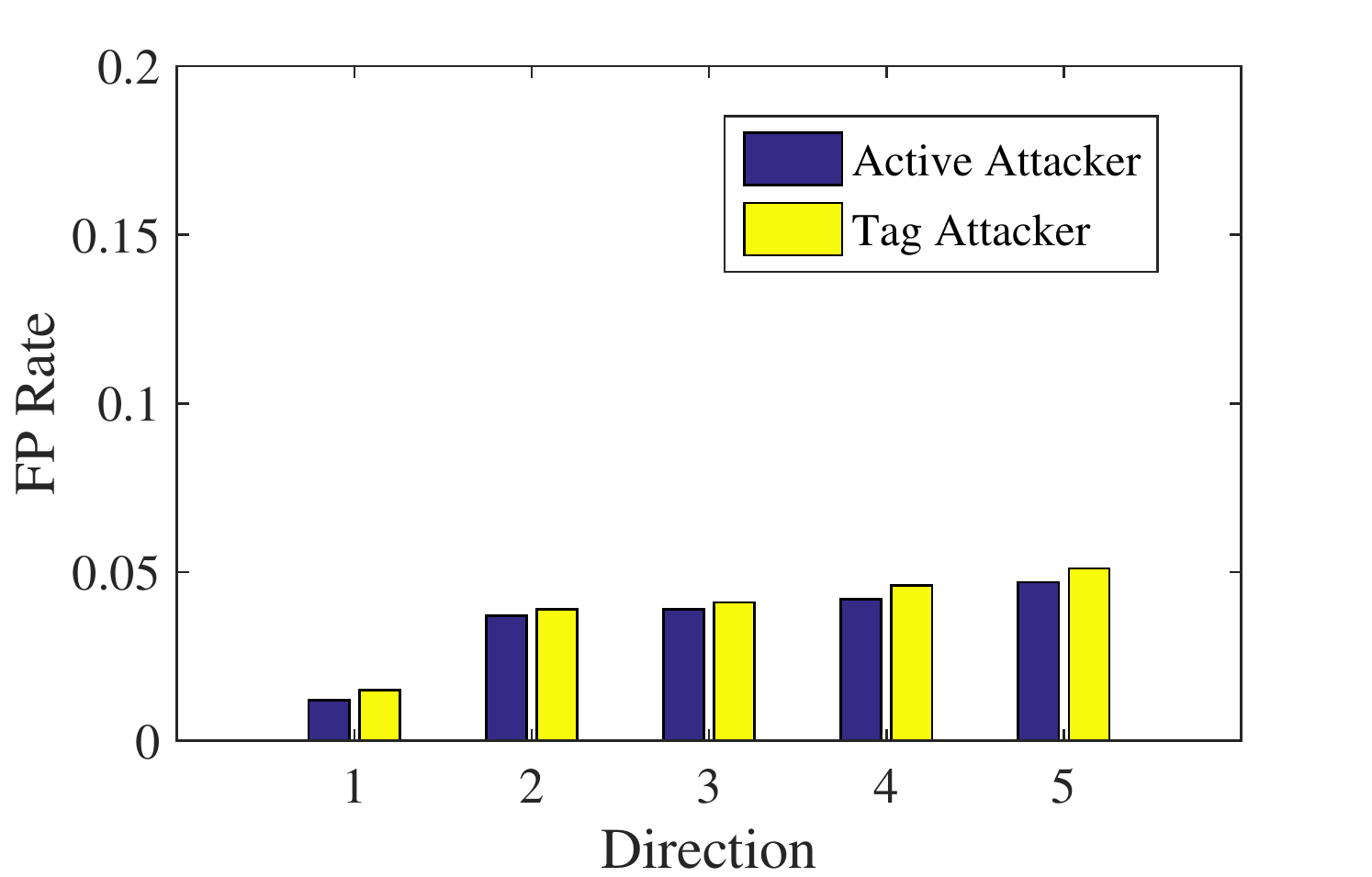}}
	\subfigure[TP rates.]
	{\includegraphics[width=2.7in]{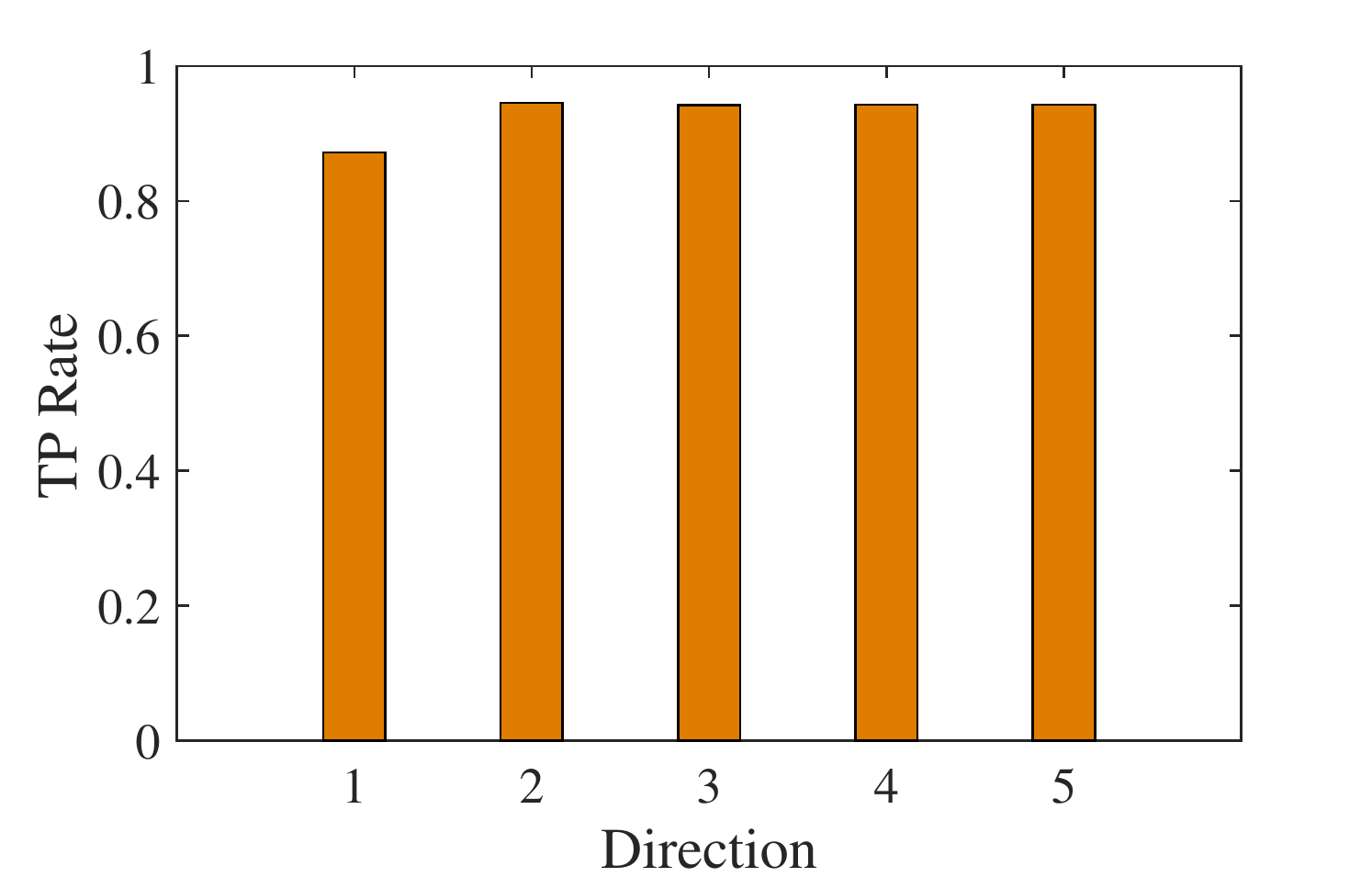}}
	\caption{TP and FP rates at different directions in static environment.}
	\label{fig:static_direction}\vspace{-0.3cm}
\end{figure}
\begin{figure}[t]
	\subfigure[FP rates.]
	{\includegraphics[width=2.7in]{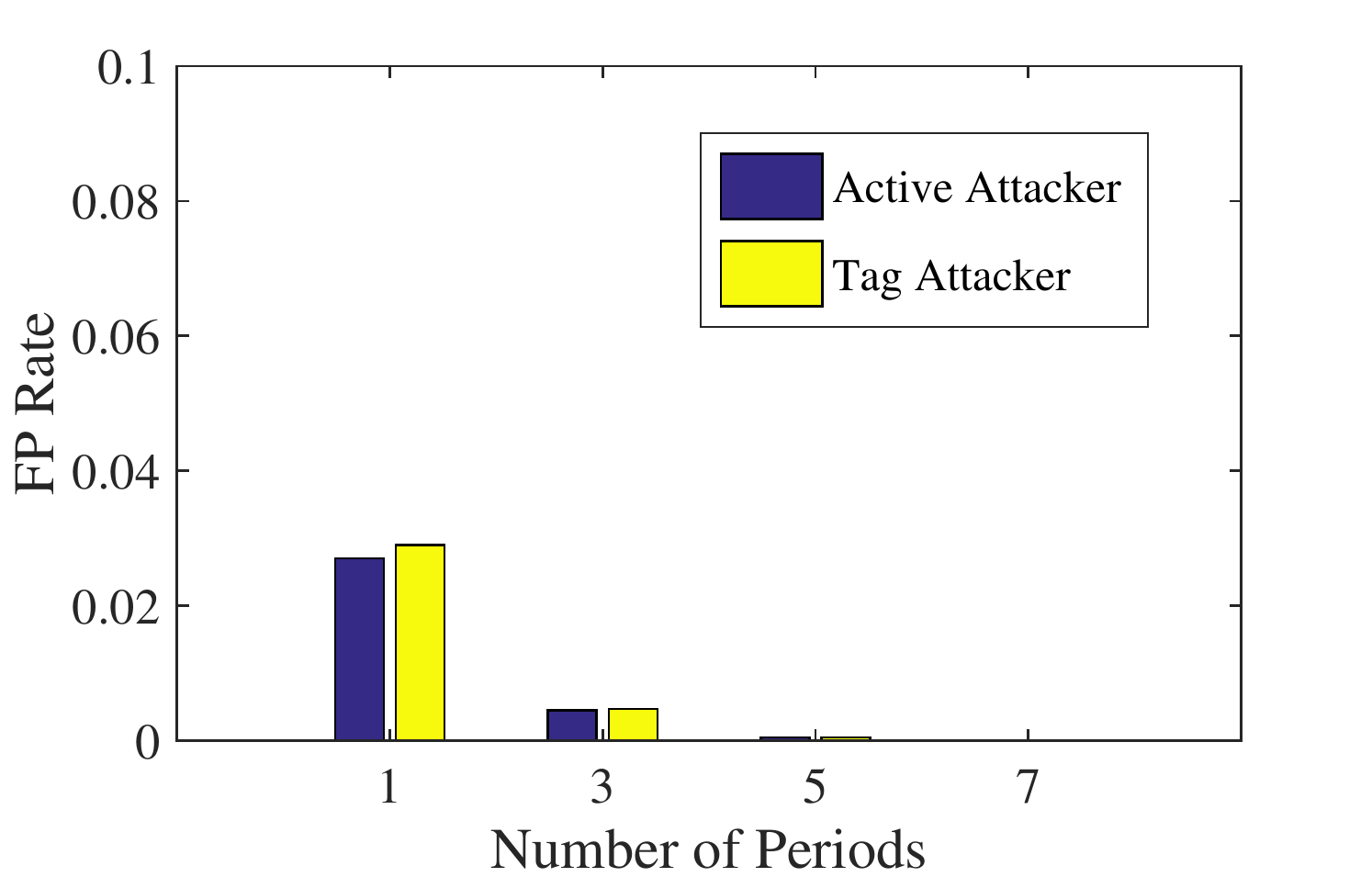}}
	\subfigure[TP rates.]
	{\includegraphics[width=2.7in]{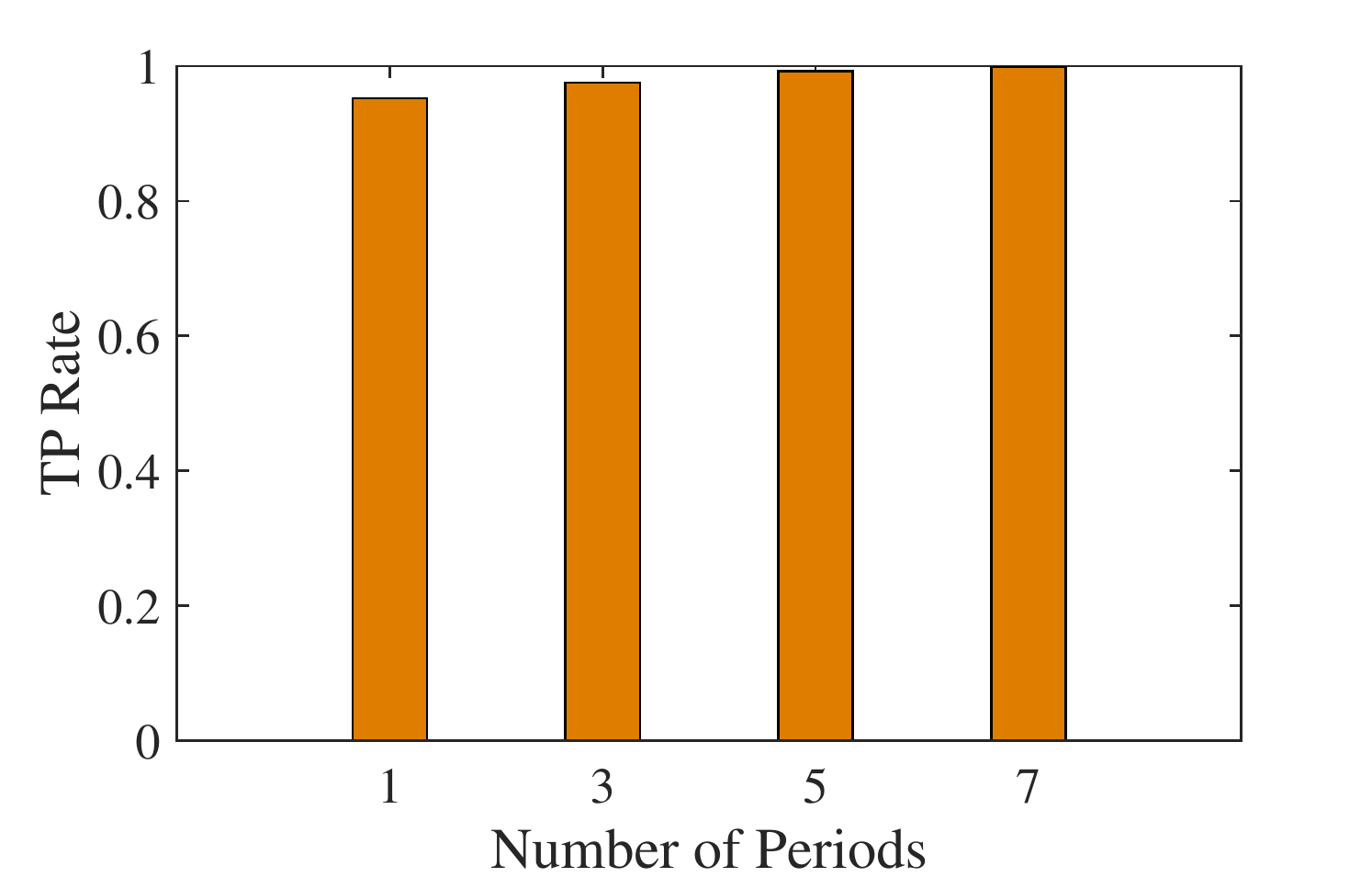}}
	%    \subfigure[Off-body RSS after fluctuation removal.]
	%    {\hspace{0.35cm}\includegraphics[width=1.65in]{figs//new/ica_plot/off_removal}}
	\caption{TP and FP rates of different times in static environment.}
	\label{fig:period} \vspace{-0.3cm}%\vspace{-0.3cm}
\end{figure}
\textbf{Results}. Fig.~\ref{fig:static_distance} shows the FP and TP rates when the tag and active attacker are at different distances from the user under the static environment. The results reveal that when the distances between the attacker and the user are less than 30~cm, the average TP rate is 94\% but the FP rate is larger than 10\%. However, when the distances are larger than 50~cm, SecureScatter can achieve very low FP rates of 0-3\% and TP rates of 94-96\%. This is because the body will have a significant effect on the multi-path of the attacker within 50~cm and the tag attacker is more easily influenced by the human body due to the reflected path. We conclude that the safe distance of SecureScatter is 50~cm, which is hard for attackers to hide them at such a short distance.

\begin{figure}[t]
    %    \center 
    \subfigure[FP rates.]
    {\includegraphics[width=2.7in]{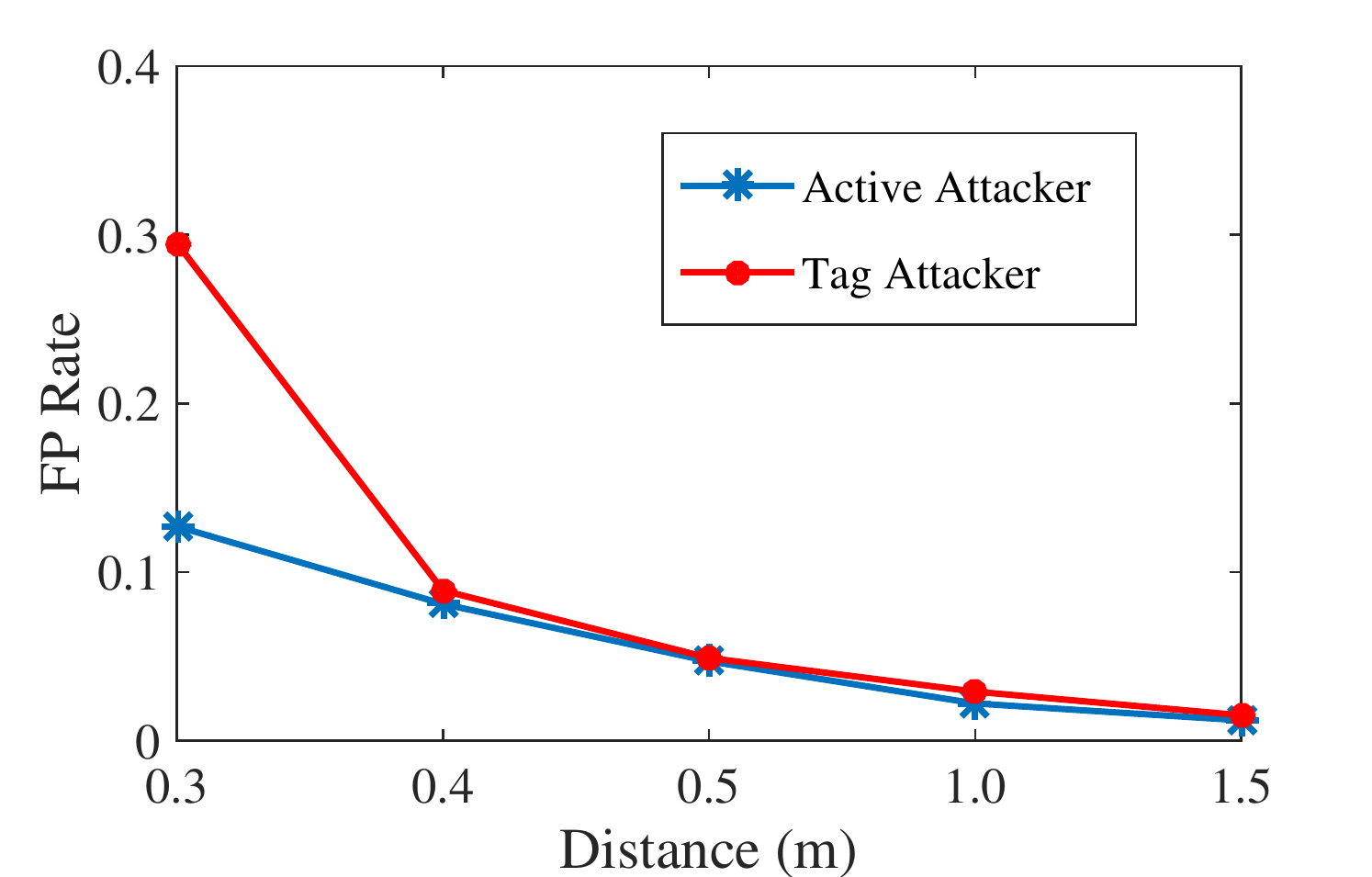}}
    \subfigure[TP rates.]
    {\includegraphics[width=2.7in]{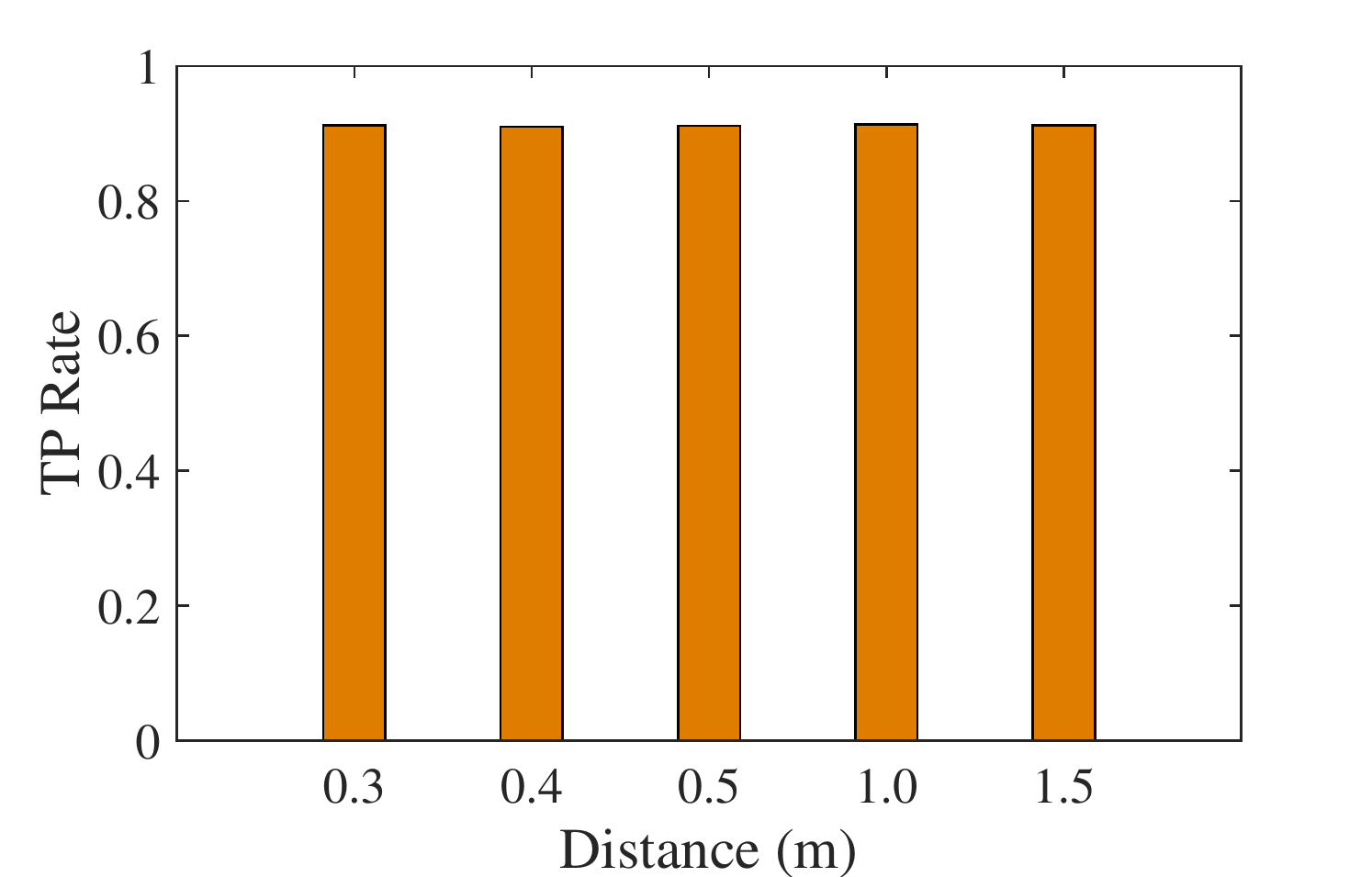}}
    \caption{TP and FP rates at different distances in dynamic environment.}
    \label{fig:mobile_distance}\vspace{-0.3cm}
\end{figure}
\begin{figure}[t]
    \subfigure[FP rates.]
    {\includegraphics[width=2.7in]{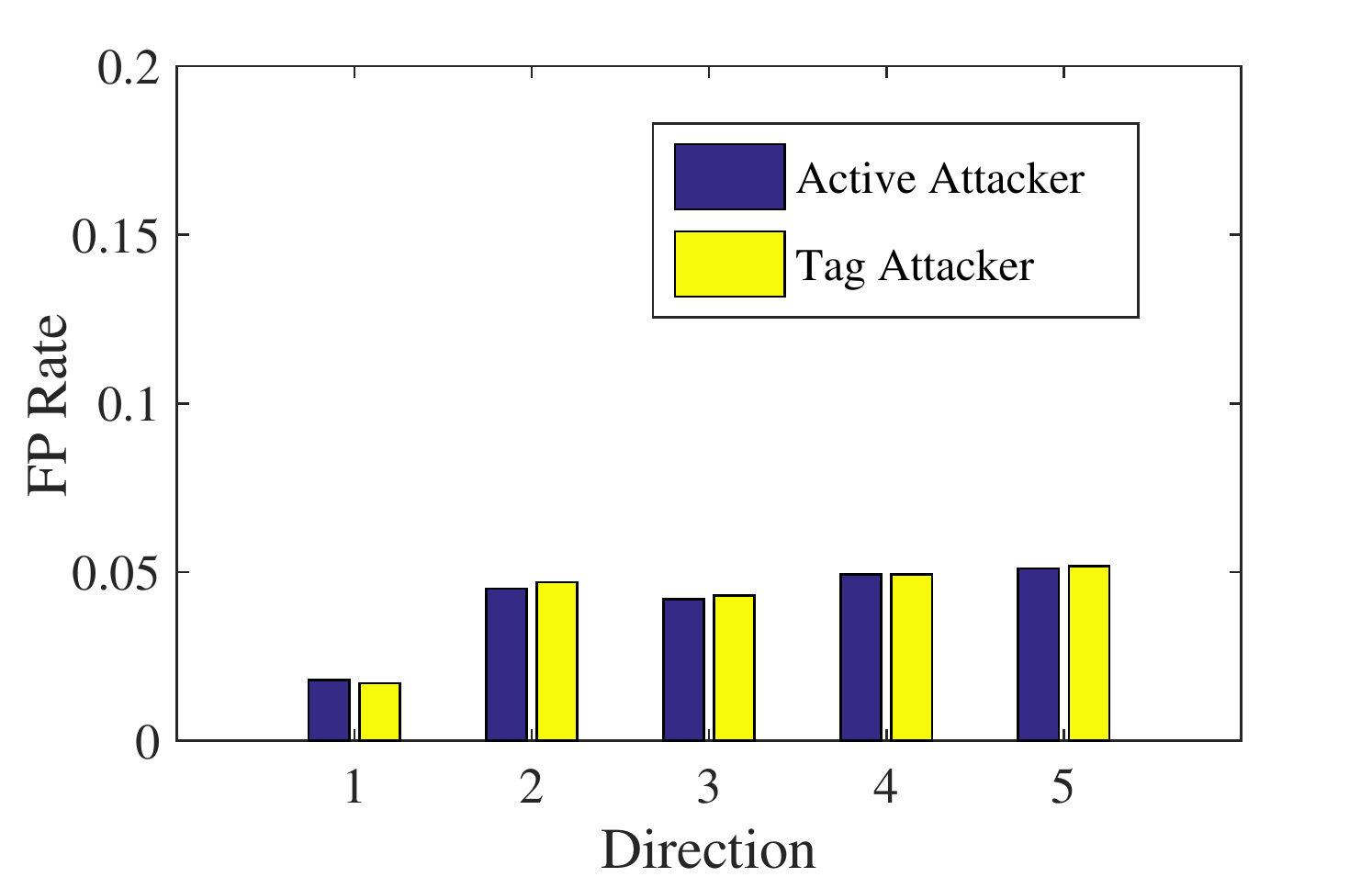}}
    \subfigure[TP rates.]
    {\includegraphics[width=2.7in]{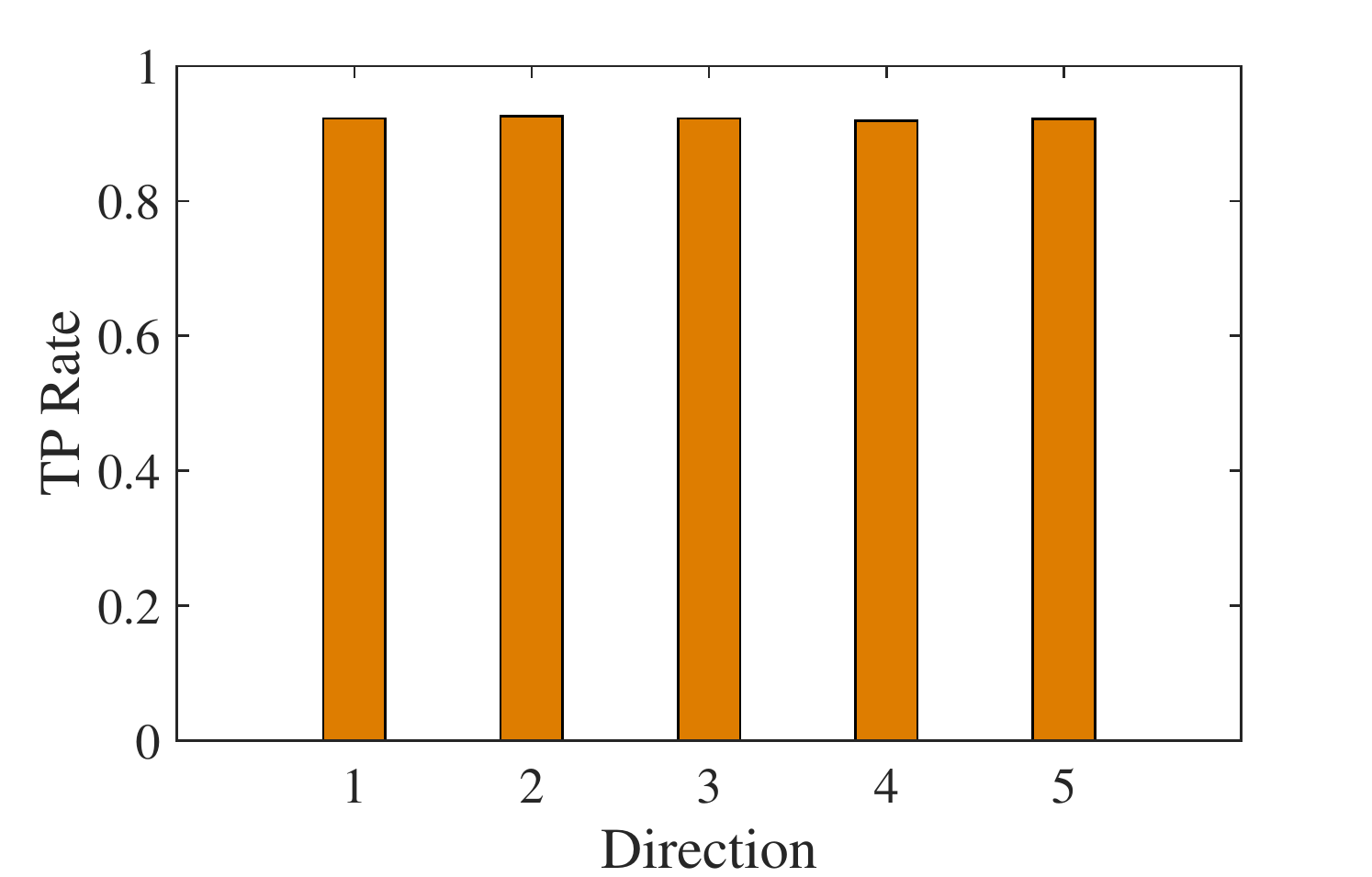}}
    \caption{TP and FP rates at different directions in dynamic environment.}
    \label{fig:mobile_direction}\vspace{-0.3cm}
\end{figure}

Fig.~\ref{fig:static_direction} evaluates the impacts of different directions of the attacker. We observe that the place 1 in direction of line-of-sight (LOS) achieves the lowest FP rates of 2\%, while the place 2 and 3 where are the obstacles near between the attacker and user achieve FP rates of 3-5\%. However, the place 4 and 5 where are aside from the body achieve the worst FP rates of 5\%. This means that the effects of body shield are larger than the fixed obstacles.

Next, we employ more number of times of the movements to enhance the reliability of the system and the results are shown in Fig.~\ref{fig:period}. When we perform the experiment just for one time, the TP rate is 95.23\%. While using more than five motions, it can boost the TP rates to over 99\%. On the other hand, with five times, we can achieve near-zero FP rates.

\subsection{Dynamic Environment}\label{sec:unstable_experiments} 
In this section, we test the performance of SecureScatter when the user is under dynamic environments, in which, we consider two types of the common states, i.e., dynamic body and dynamic surrounding around the user. For the dynamic body state, the user is asked to finish some slight motions such as hand movement, body movement or head shake. As for the dynamic surrounding around, we need two people to walk around the user when testing. Then we evaluate the performance of SecureScatter based on the distances and directions following the same settings for on-body devices and attacker as the first set of the experiments under static environments.

\textbf{Dynamic body.} In this experiment, we request the user to perform slight motions, such as the movements of the hand, head, and other body parts. Then we evaluate the performances when attacker is at various distances and directions.

Fig.~\ref{fig:mobile_distance} and Fig.~\ref{fig:mobile_direction} show the FP and TP rates with attackers at various distances and directions under the dynamic body conditions. Fig.~\ref{fig:mobile_distance} indicates that even though there are slight motions, SecureScatter still yields an average TP rate of 92\% and FP rate lower than 5\% when the distances between the user and attacker are larger than 50~cm. We achieve relatively the same performance at different directions, which implies that our system is robust to the dynamic body states. 
\begin{figure}[t]
    \subfigure[FP rates.]
    {\includegraphics[width=2.7in]{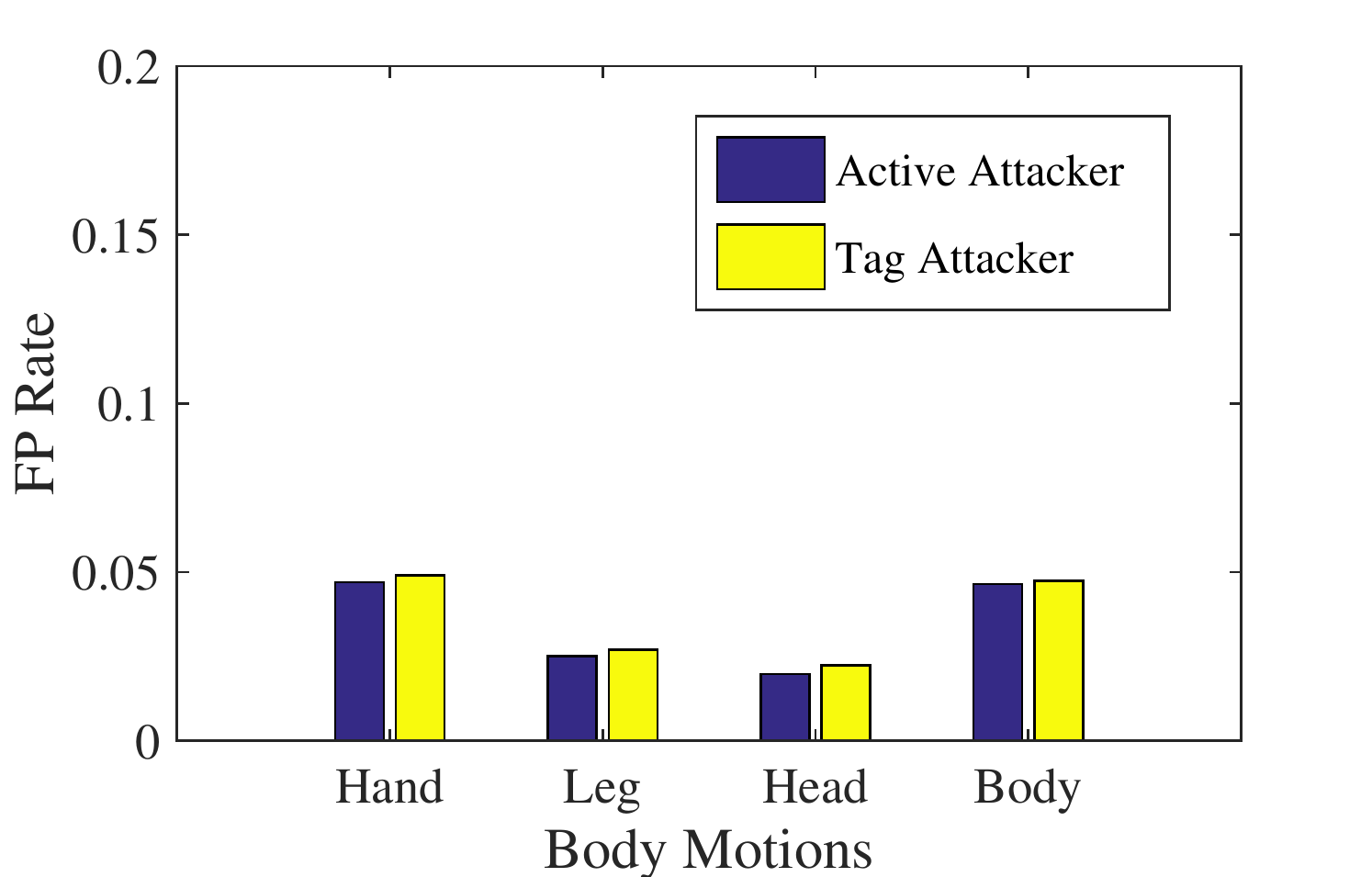}}
    \subfigure[TP rates.]
    {\includegraphics[width=2.7in]{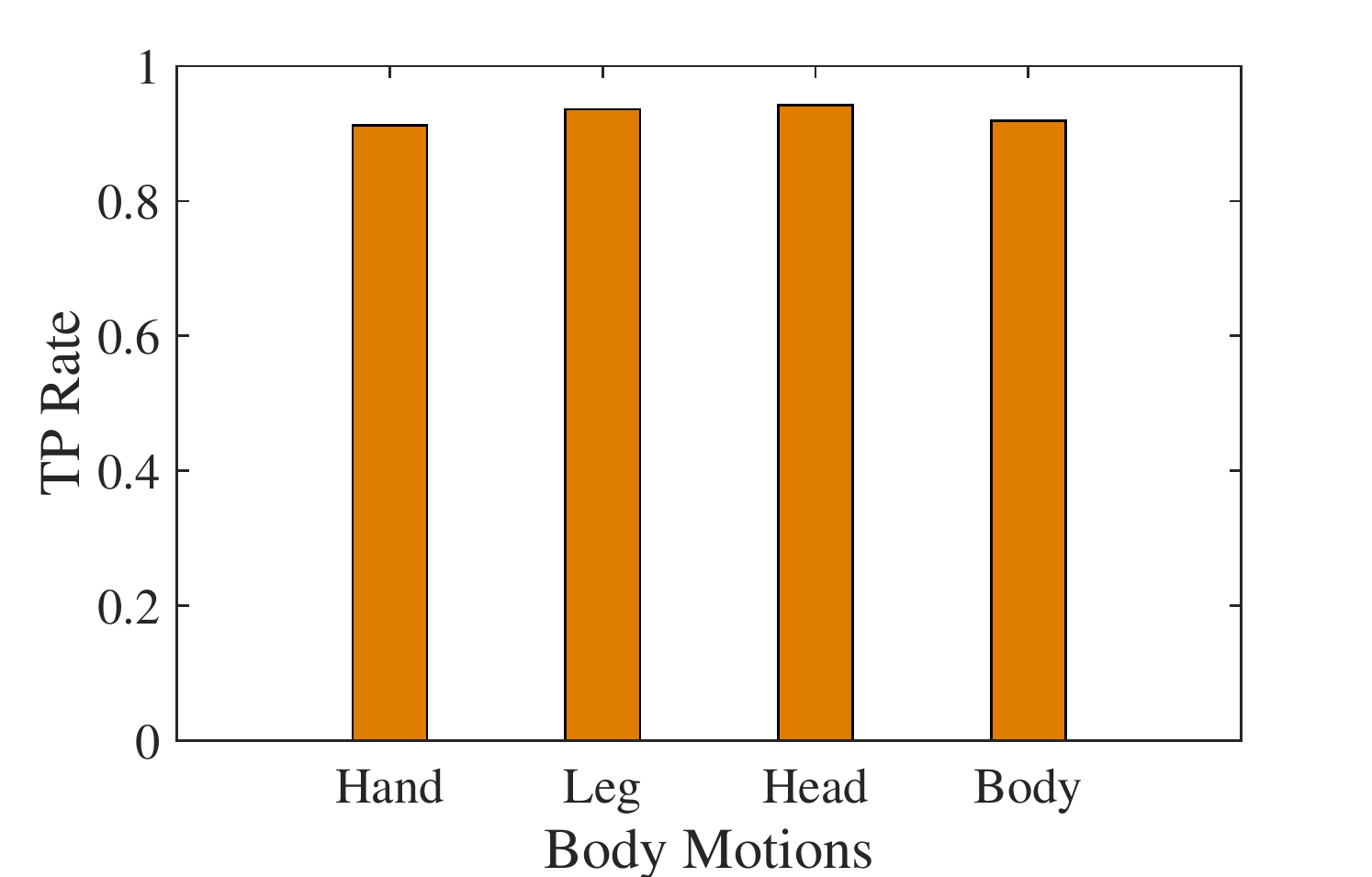}}
    \caption{TP and FP rates of the body movements.}
    \label{fig:body_motion}\vspace{-0.3cm}
\end{figure}
\begin{figure}[t]
    \subfigure[FP rates.]
    {\includegraphics[width=2.7in]{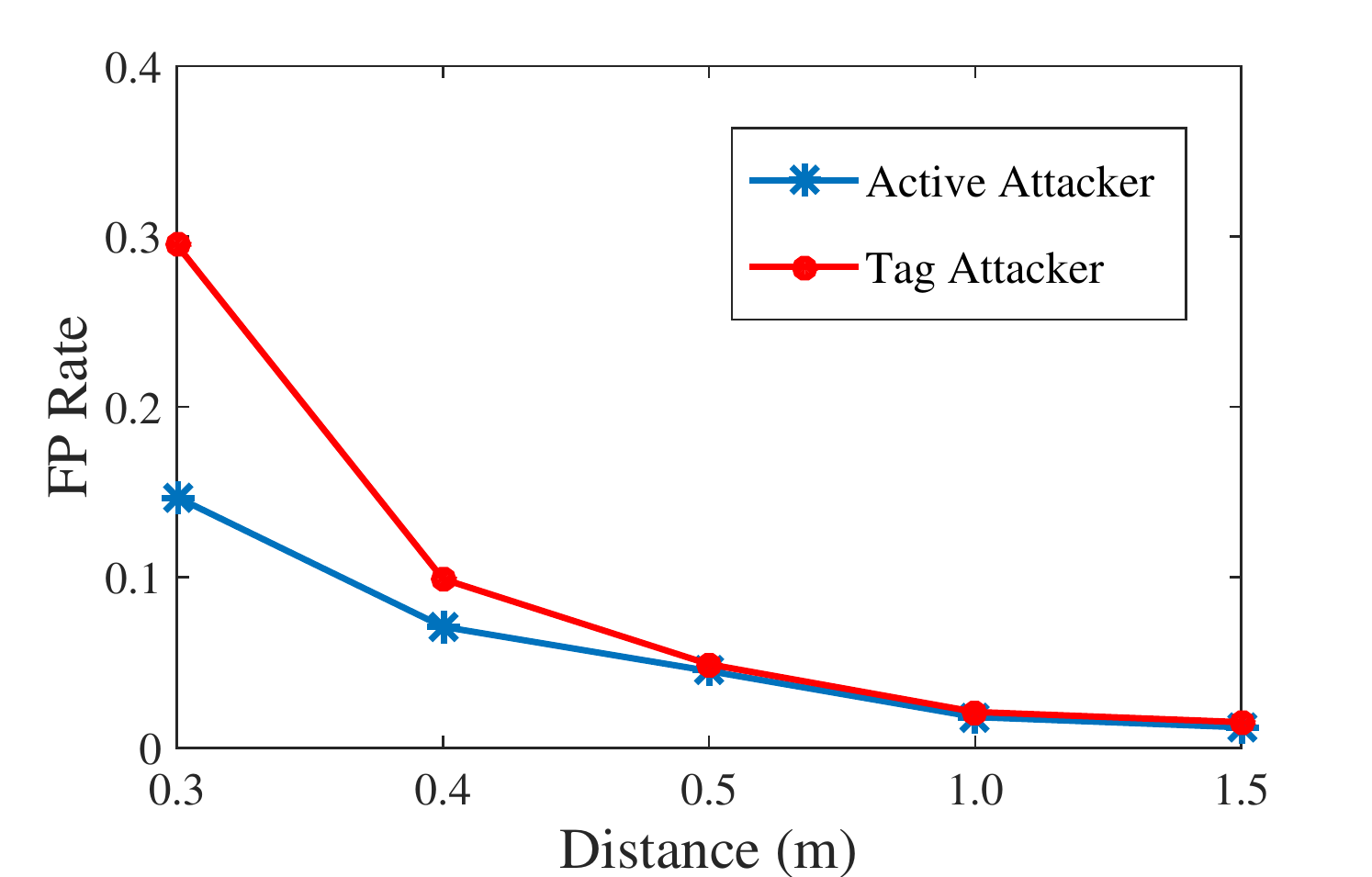}}
    \subfigure[TP rates.]
    {\hspace{0.1cm}\includegraphics[width=2.7in]{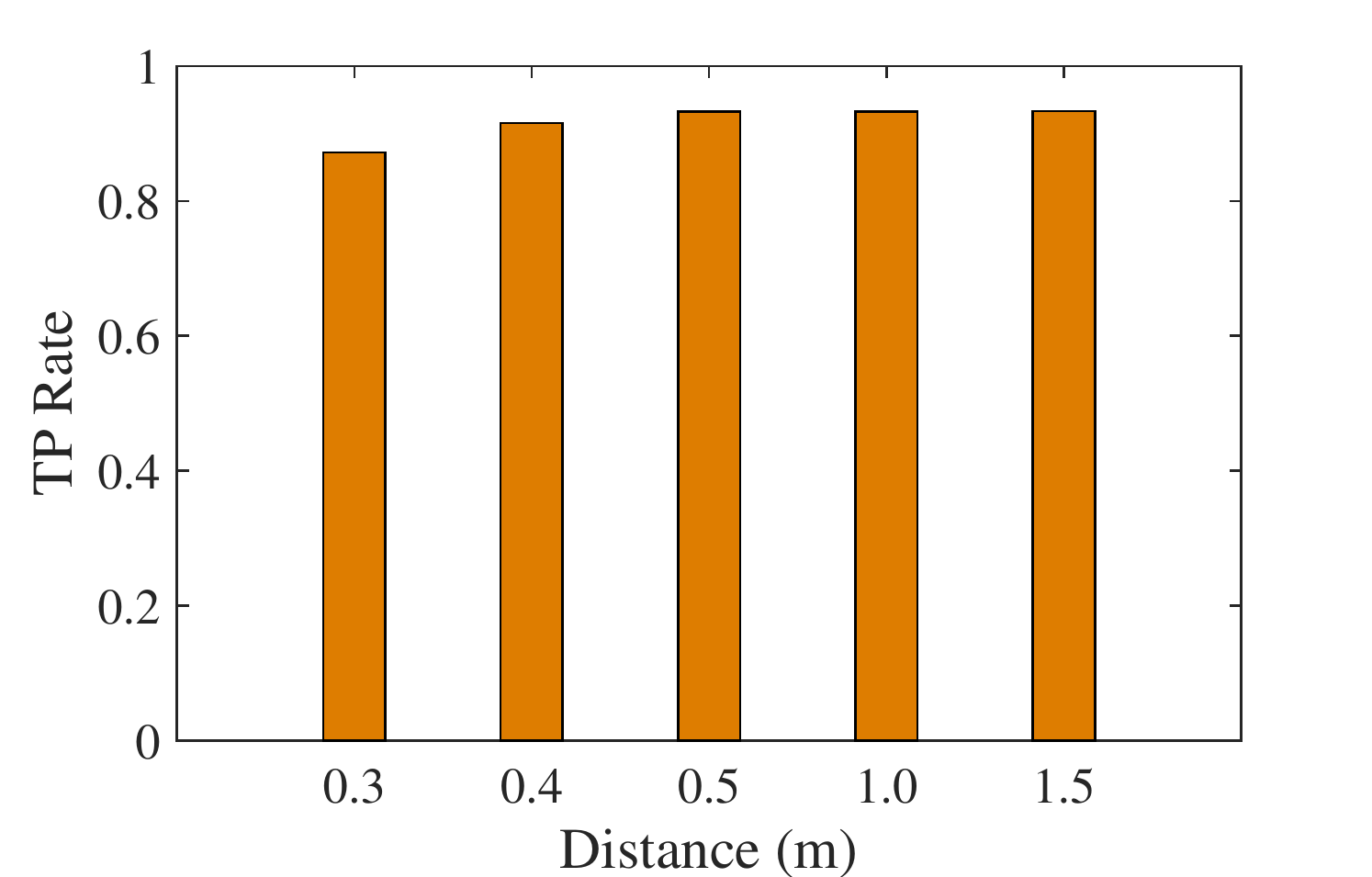}}
    
    %    \subfigure[Off-body RSS after fluctuation removal.]
    %    {\hspace{0.35cm}\includegraphics[width=1.65in]{figs//new/ica_plot/off_removal}}
    \caption{TP and FP rates when people around}
    \label{fig:body_near}\vspace{-0.3cm}
\end{figure}

Fig.~\ref{fig:body_motion} shows the FP and TP rates under different body movements. We observe that even though signal variations are affected by different body movements, SecureScatter can still achieve the average TP rate higher than 92.3\% and FP rate lower than 5\%. This implies that our system is robust against different body movements. Additionally, by comparing TP rates of the different motions, we observe that the rates of the motions from the leg movements and head movements are larger than that under the body movements and hand movements. This is because the on-body tag is more sensitive to the nearby motions.

\textbf{Dynamic surroundings.} Dynamic surroundings such as walking people affect the radio propagations. In this experiment, we evaluate the robustness of SecureScatter in such dynamic surroundings by measuring the FP and TP rates when the people walk around at different distances from the user.

Fig.~\ref{fig:body_near} shows that when the distances between the user and people around are larger than 50~cm, SecureScatter can achieve the FP rate as low as below 5\% and TP rate larger than 95\%, which indicates that SecureScatter yields a reliable performance even in presence of environmental dynamics caused by walking people. On the other hand, when the distances are less than 50~cm, the FP rate is larger than 5\%. This is because the movements that are close to the user lead to variations in the paths reflected by the moving body, which are similar to on-body backscatter paths and thereby disguise the behaviors of attackers.
\begin{figure}[t]
	\subfigure[FP rates.]
	{\includegraphics[width=2.7in]{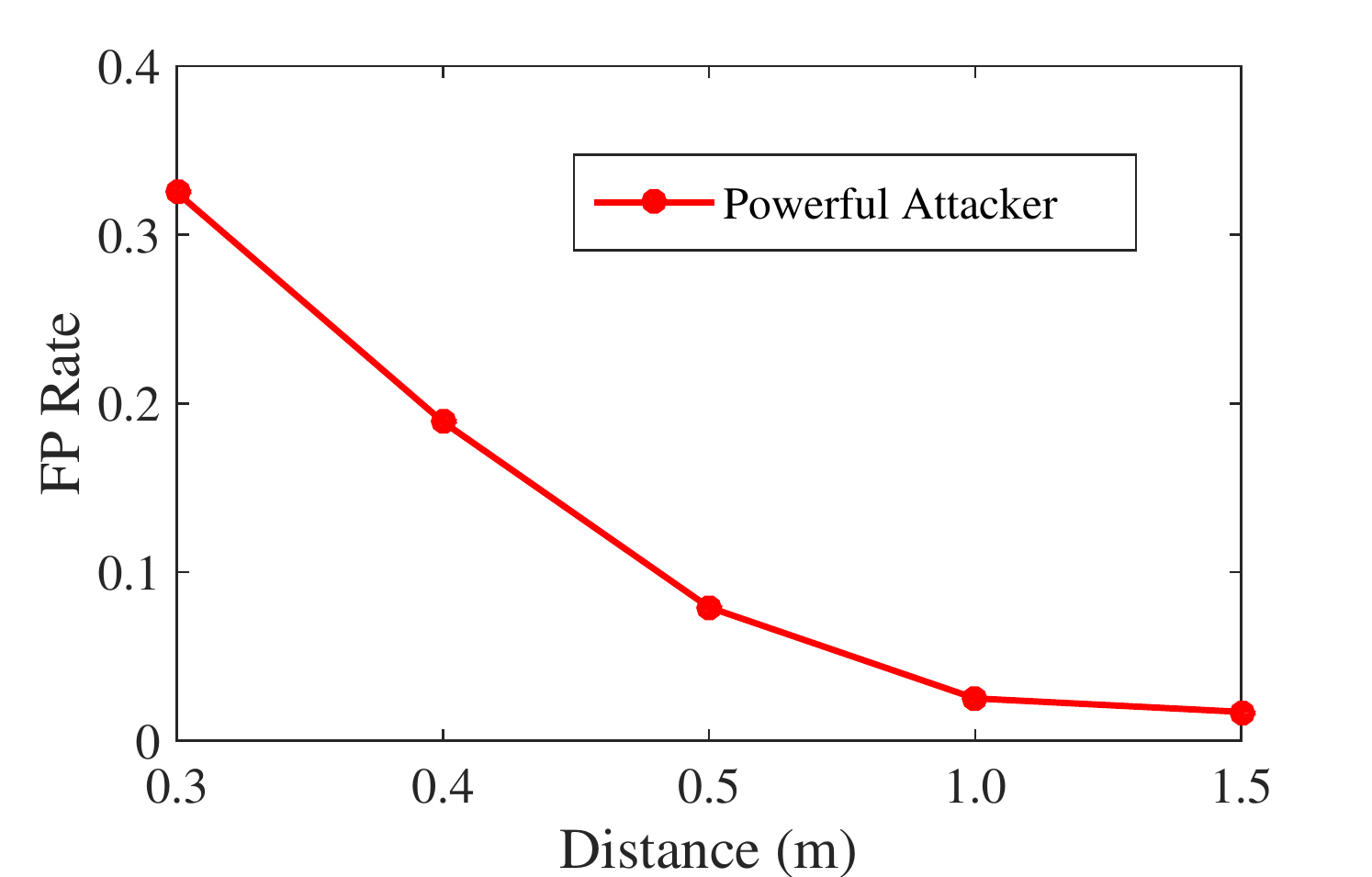}}
	\subfigure[TP rates.]
	{\hspace{0.1cm}\includegraphics[width=2.7in]{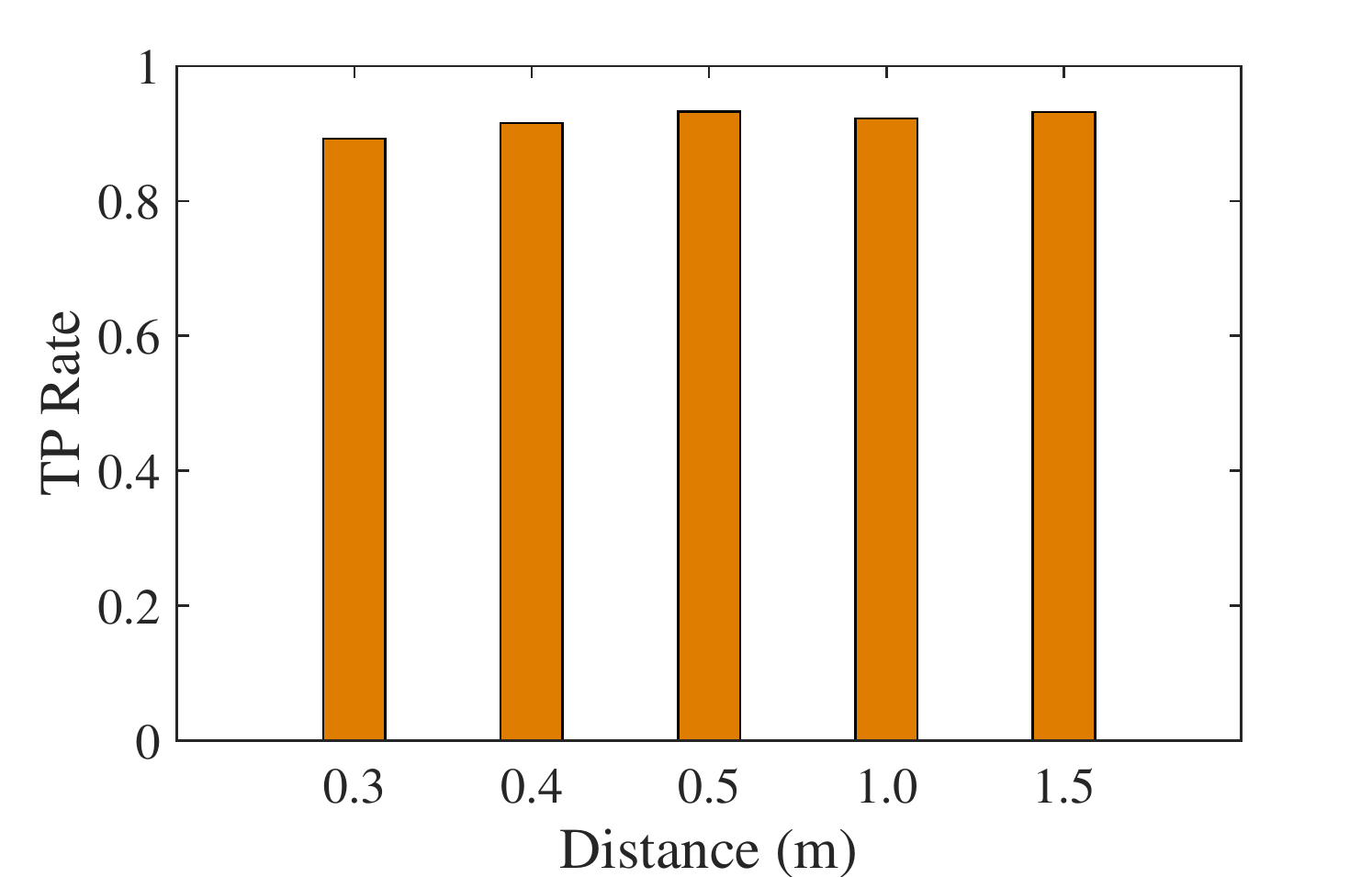}}
	\caption{TP and FP rates when attacked by powerful active attacker}
	\label{fig:powerful_attacker}\vspace{-0.3cm}
\end{figure}

\begin{figure}[t]
	%    \center 
	\subfigure[FP rates.]
	{\includegraphics[width=2.7in]{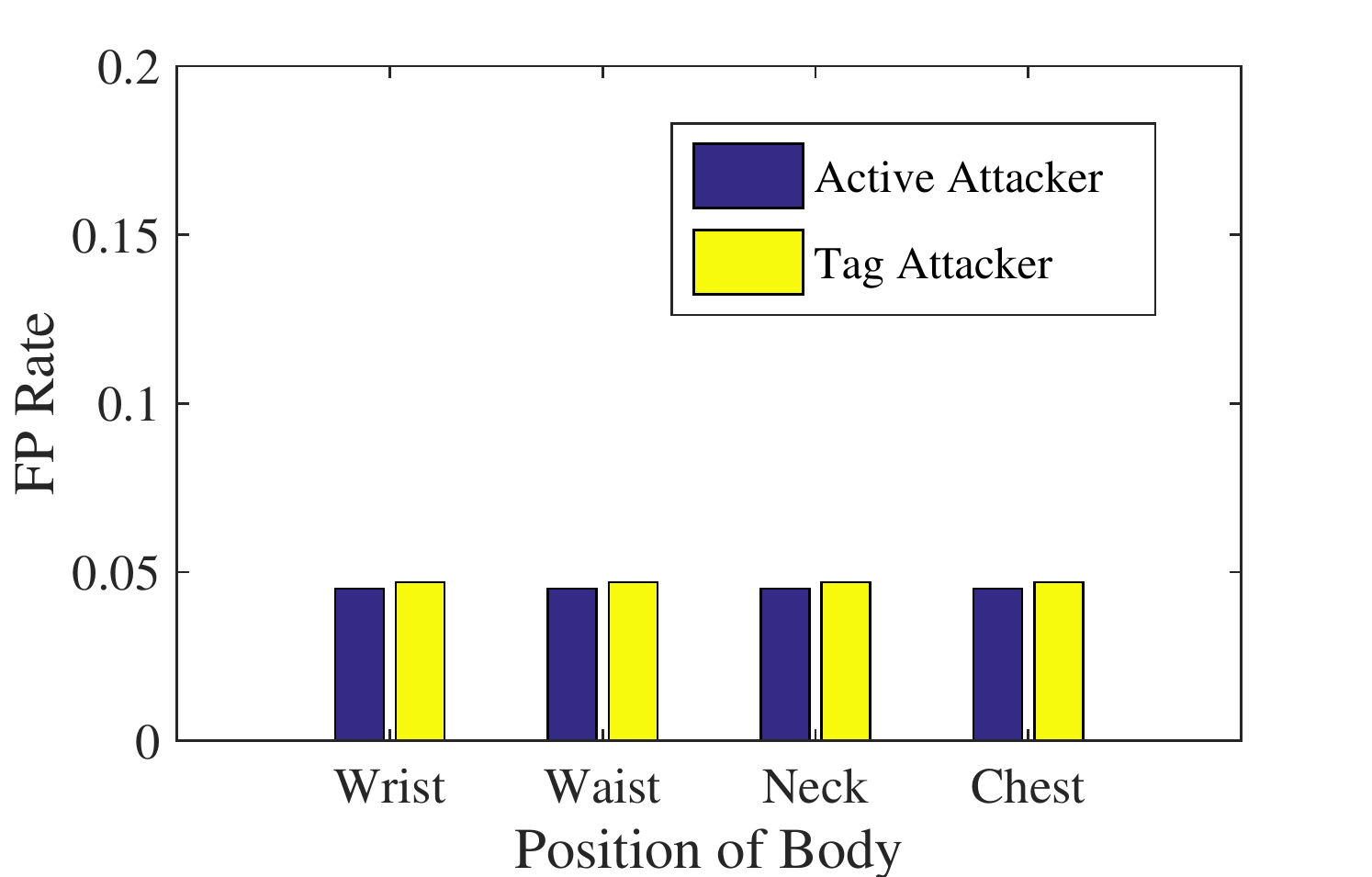}}
	\subfigure[TP rates.]
	{\hspace{0.1cm}\includegraphics[width=2.7in]{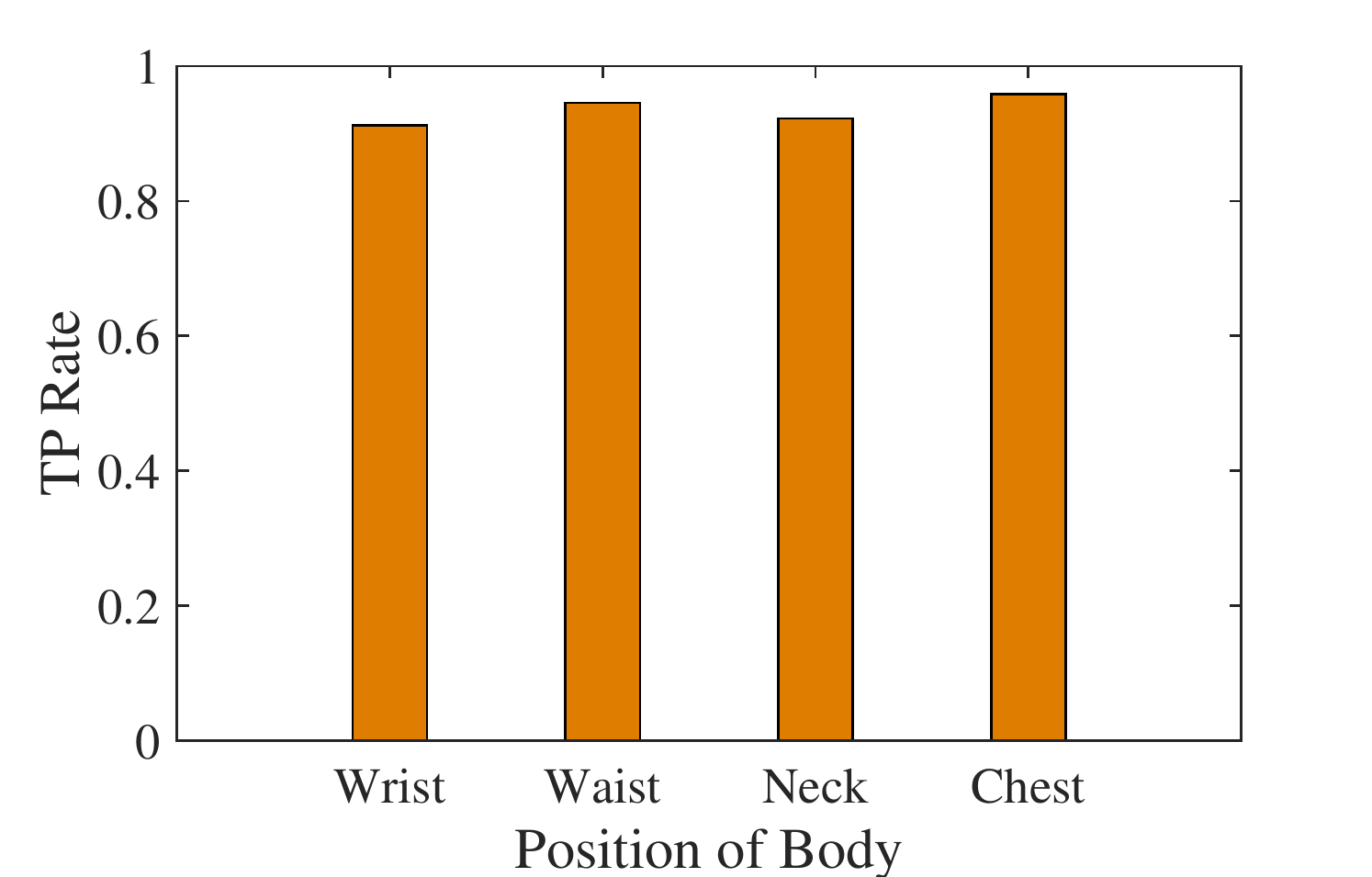}}
	\caption{TP and FP rates in different positions of body.}
	\label{fig:body_position}\vspace{-0.3cm}
\end{figure}

% \begin{figure}[t]
%%     \center
%     \includegraphics[width=1.7in]{body_position.pdf}\vspace{-0.2cm}
%     \caption{cluster}
%     \label{fig:cluster}\vspace{-0.3cm}
%    \end{figure}
%     \begin{figure}[t]
%%         \center
%         \includegraphics[width=1.7in]{body_position.pdf}\vspace{-0.2cm}
%         \caption{cluster}
%         \label{fig:cluster}\vspace{-0.3cm}
%        \end{figure}
%\begin{figure}[t]
%    %    \center 
%    \subfigure[FP rate of sta-distance brokenline.]
%    {\includegraphics[width=1.7in]{mobile_direction.pdf}}
%    \subfigure[FP body position.]
%    {\hspace{0.1cm}\includegraphics[width=1.7in]{body_position.pdf}}
%    
%    %    \subfigure[Off-body RSS after fluctuation removal.]
%    %    {\hspace{0.35cm}\includegraphics[width=1.65in]{figs//new/ica_plot/off_removal}}
%    \caption{Illustration of signal decomposition.}
%    \label{fig:ica_plot}\vspace{-0.3cm}
%\end{figure}
\subsection{Powerful Active Attacker}\label{sec:powerful attack} 
We consider the effects of a powerful active attacker that can transmit varying powers to mislead the on-body receiver by monitoring the RSS variations. In our experiment, we employ a USRP as the powerful active attacker. In order to emulate RSS variations monitoring, we use two antennas, one of which is used to monitor the RSS and the other is used to transmit the fake data. If the monitoring RSS variation is larger than 4~dB, the attacker will consider the on-body transmitter is performing authentication movement and then change the power of the fake data so as to follow the movement of the on-body transmitter. After that, the on-body receiver receives the signal and performs authentication.

\begin{figure}[t]
	\subfigure[FP rates.]
	{\includegraphics[width=2.72in]{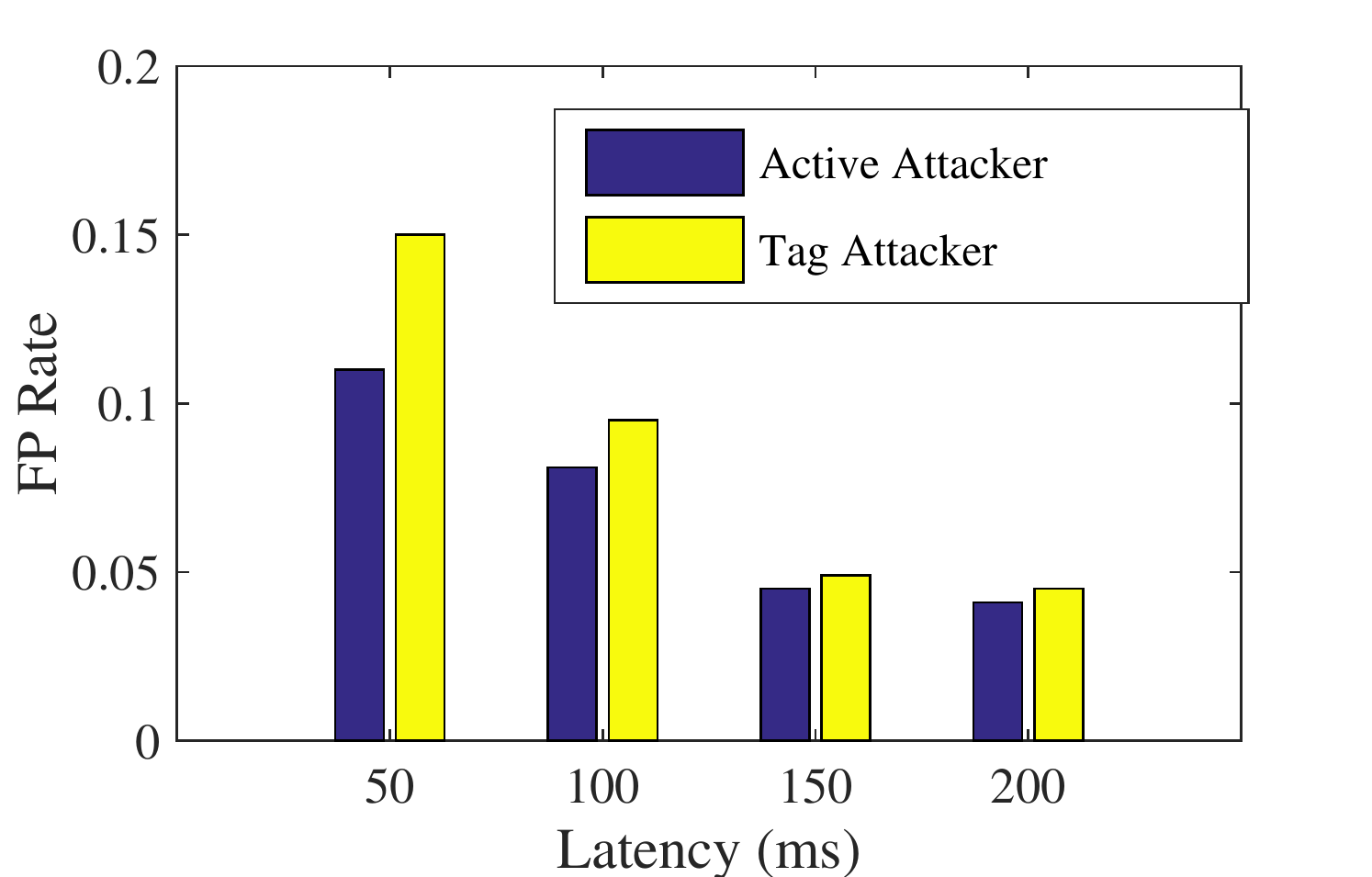}} 
	\subfigure[TP rates.]
	{\includegraphics[width=2.72in]{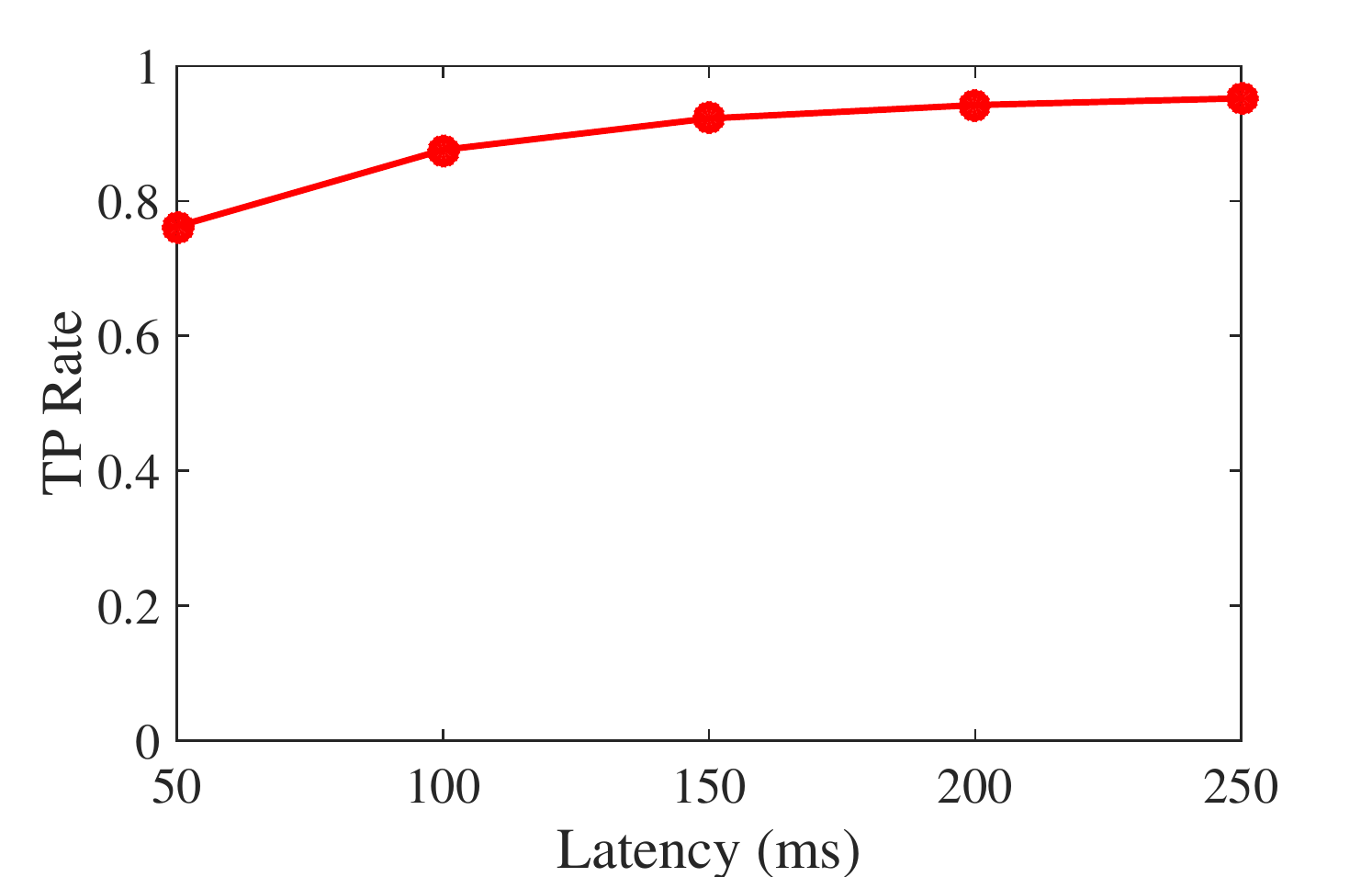}}
	%    \subfigure[Off-body RSS after fluctuation removal.]
	% {\hspace{0.35cm}\includegraphics[width=1.65in]{figs//new/ica_plot/off_removal}}
	\caption{TP and FP rates of different latency.}
	\label{fig:latency}\vspace{-0.3cm}
\end{figure}
\begin{figure}[t]
	%    \center 
	\subfigure[FP rates.]
	{\includegraphics[width=2.7in]{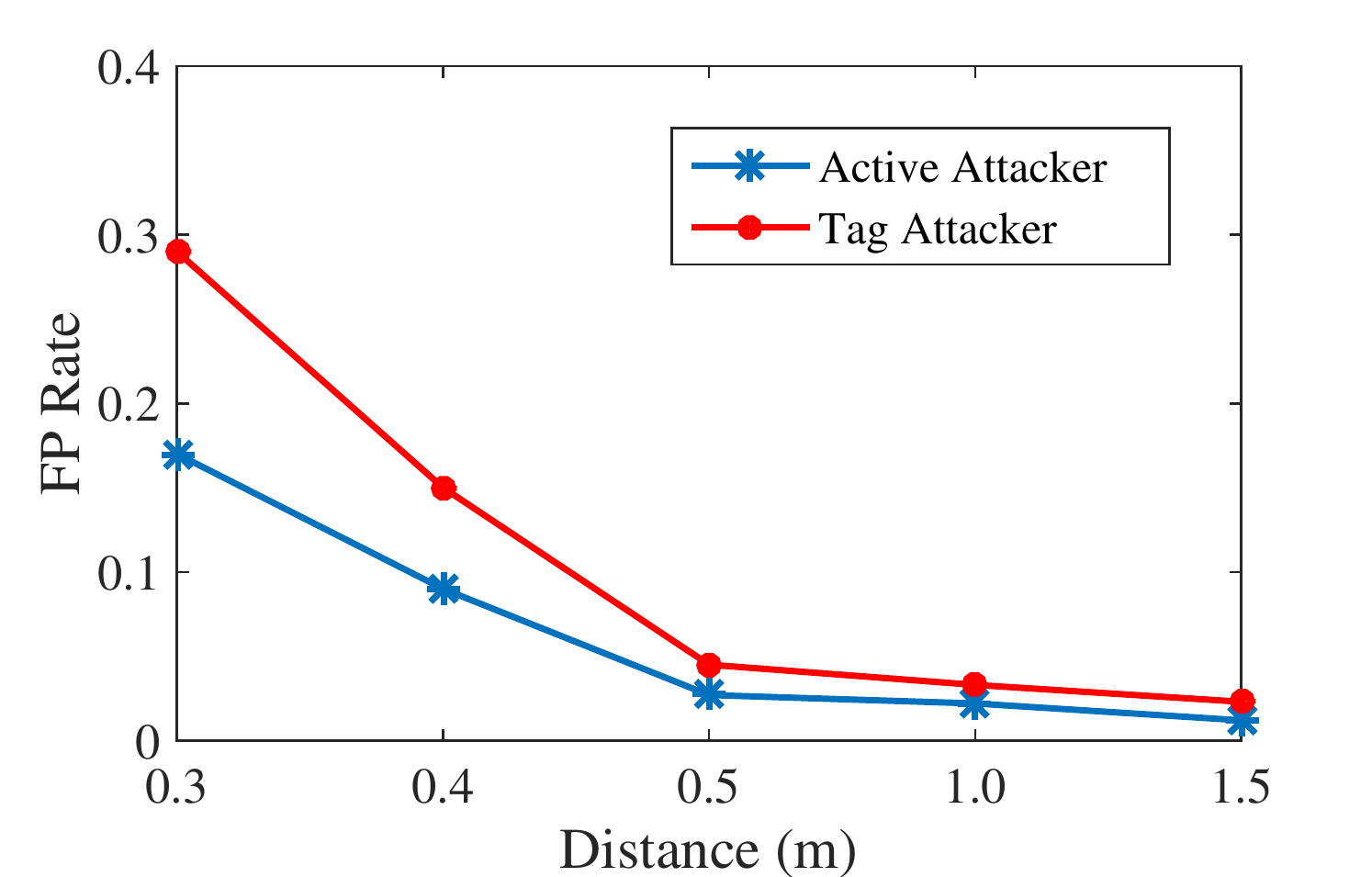}}
	\subfigure[TP rates.]
	{\includegraphics[width=2.7in]{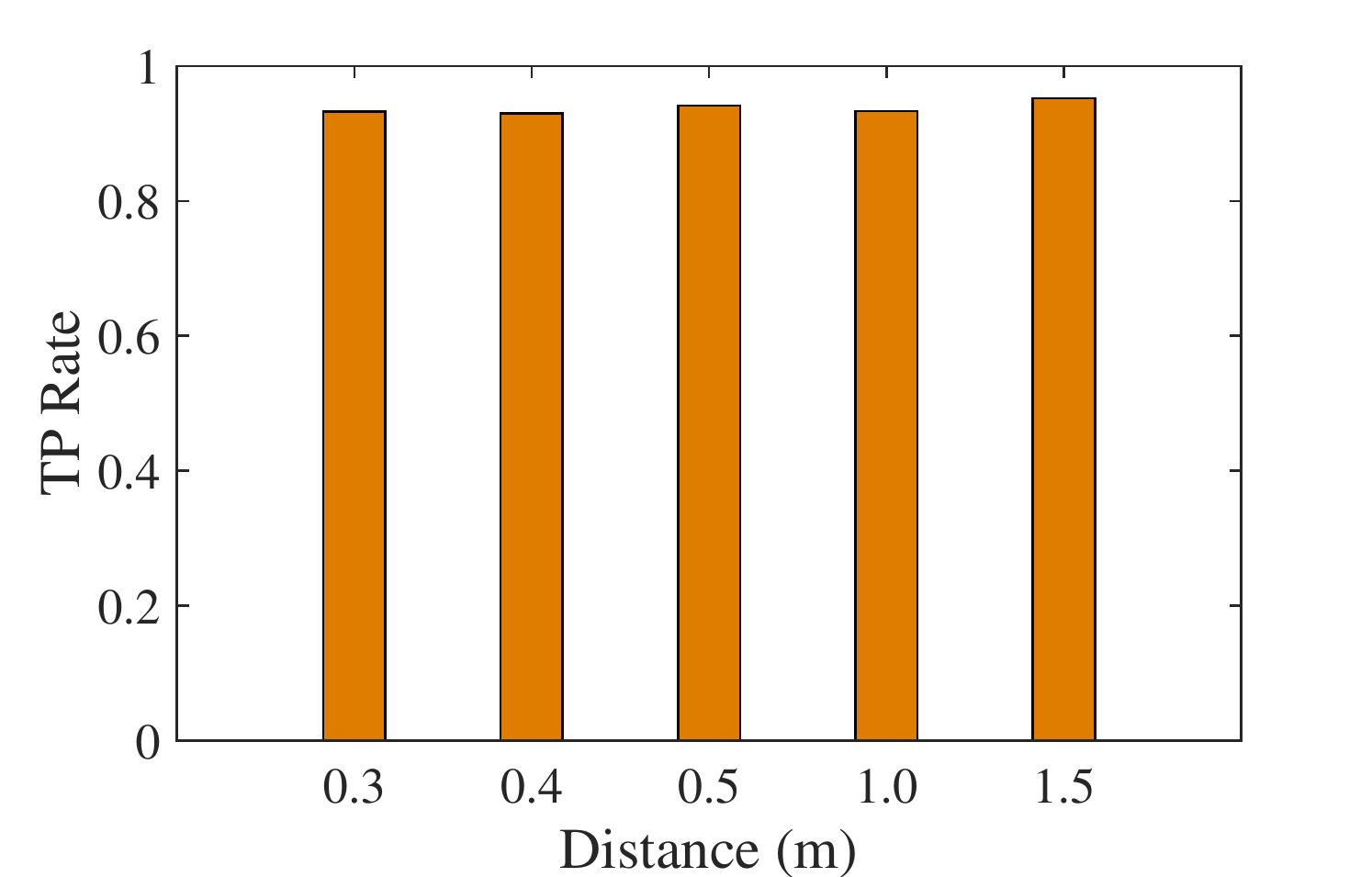}}
	%    \subfigure[Off-body RSS after fluctuation removal.]
	% {\hspace{0.35cm}\includegraphics[width=1.65in]{figs//new/ica_plot/off_removal}}
	\caption{TP and FP rates at different distances at 2.4~GHz.}
	\label{fig:wifi}\vspace{-0.3cm}
\end{figure}
Fig.~\ref{fig:powerful_attacker} presents the FP and TP rates of the system attacked by the powerful attackers. Based on the result, we can see that FP rates in different distances of the body are different. When the attacker is far away from the body, the FP rates are lower than that in the short distances that because when the powerful attacker is close to the body, it can detect the RSS variations correctly and change the varying power accurately. However, when the powerful attacker is far away from the body, it is not sensitive to the RSS variations and SecureScatter can achieve an average FP rate lower than 7\% and an average TP rate of 92\%.

\subsection{Wearing Position}\label{sec:overall} 
Different positions on the body can also affect the radio propagation of the on-body devices. In a practical environment, the on-body tags may be used in different positions of the body. Thus, in this subsection, we consider the performance that the tag on different positions of the body. In this experiment, we select four typical positions for the tag, i.e., the chest, waist, the wrist, and the neck, and we place the receiver on the waist and transmitter in hand. 

Fig.~\ref{fig:body_position} presents the FP and TP rates of the system in different positions of the tested user. Based on the result, we can find that TP rates in different positions of the body vary. When the tag is on the chest and waist, the TP rates are larger than that on the wrist and neck. However, even the different positions have effects on the radio propagations on the body, SecureScatter can still achieve the average TP rate of 94.23\% and FP rate lower than 5\%. It indicates that SecureScatter is robust in terms of wearing positions, which is important to be used in the practical environments.

\subsection{System Latency}\label{sec:latency} 
System latency is an important characteristic of the authentication system. As SecureScatter authenticates the on-body tag via matching the different segments and then fetching the signatures information from them. Thus, in this experiment, in order to consider the system latency, we leverage the length of the segments that the tag authentication needed to indicate the system latency.

Fig.~\ref{fig:latency} shows that when we exploit more samples to authenticate the tag, the TP rates increase at the cost of larger latencies. According to the results, when the latency is larger than 150~ms, SecureScatter can achieve TP rates as high as 92\% and FP rates lower than 4.7\%. Therefore, we can select the samples at 150~ms which is acceptable for the users.

\subsection{Working at 2.4 ~GHz}\label{sec:dec} 
We also evaluate the performance of our system at 2.4~GHz. In this experiment, firstly, we make the on-body devices work at 2.4~GHz, and then employ a tag working at 2.4~GHz to emulate the off-body tag attacker, and a USRP B210 node to emulate the active attacker. Then we evaluate the performance combined with the distances.

Fig.~\ref{fig:wifi} show the FP and TP rates. When the distances between the user and attacker are larger than 50~cm, we can achieve FP rates as low as below 5.2\%, and the average TP rate of 93.31\%. Combining the results, we can conclude that SecureScatter is robustness at the frequency of 2.4~GHz.

\begin{figure}[t]
	\subfigure[FP rates.]
	{\includegraphics[width=2.72in]{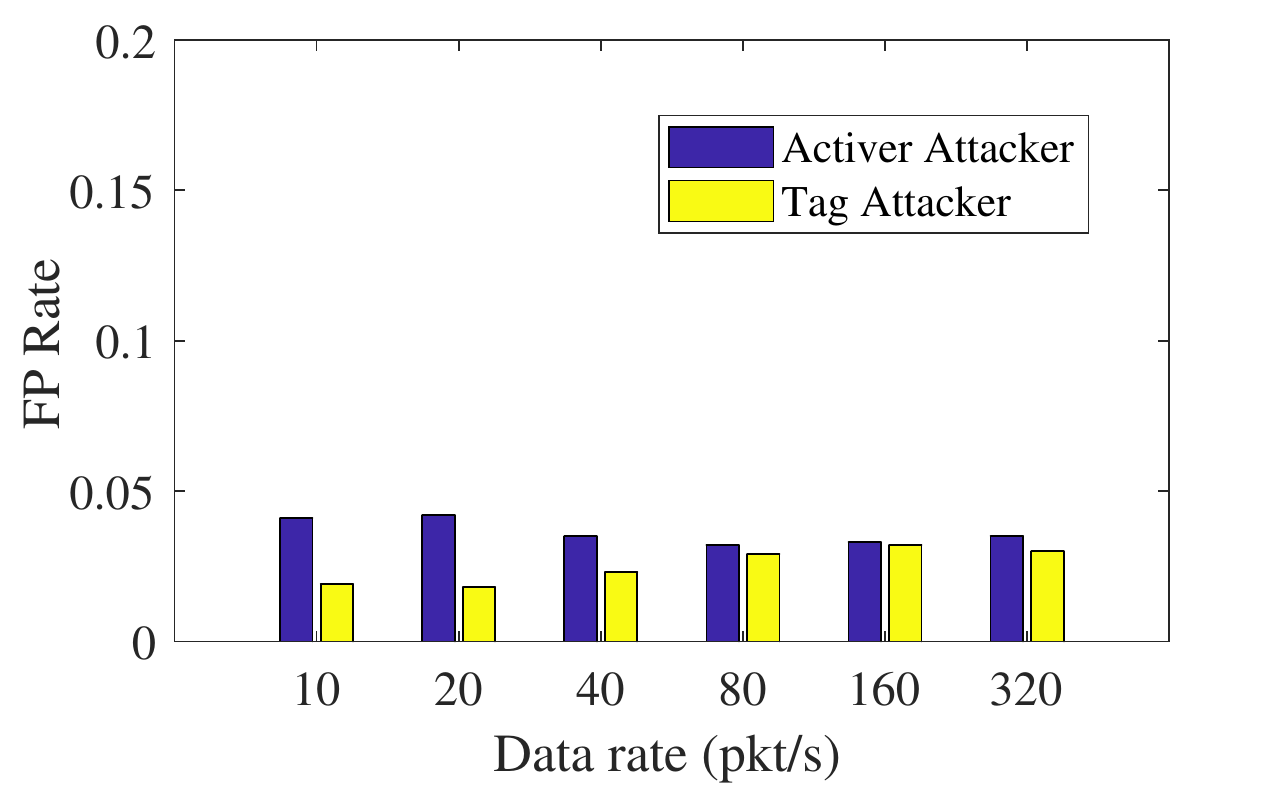}} 
	\subfigure[TP rates.]
	{\includegraphics[width=2.72in]{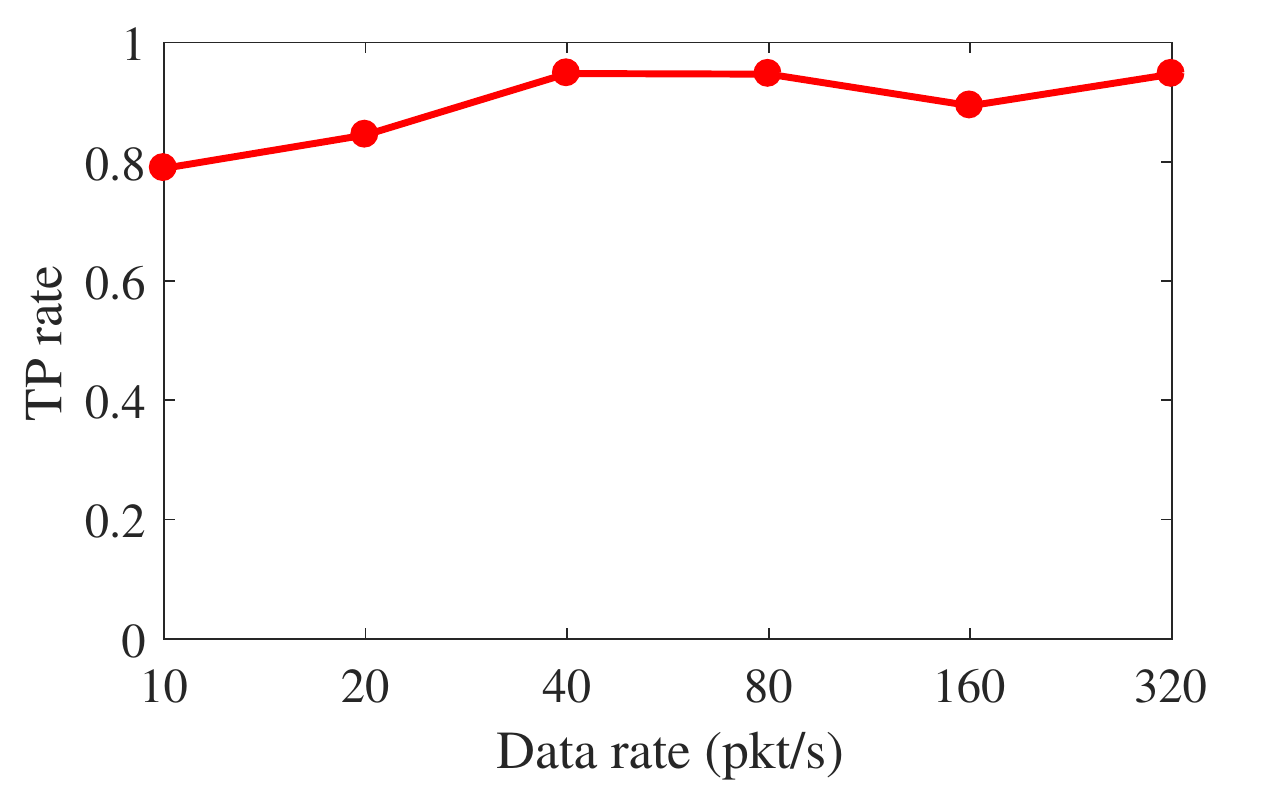}}
	%    \subfigure[Off-body RSS after fluctuation removal.]
	% {\hspace{0.35cm}\includegraphics[width=1.65in]{figs//new/ica_plot/off_removal}}
	\caption{TP and FP rates of working at non-continuous traffic.}
	\label{fig:non-con}\vspace{-0.3cm}
\end{figure}
\begin{figure}[t]
	%    \center 
	\subfigure[FP rates.]
	{\includegraphics[width=2.7in]{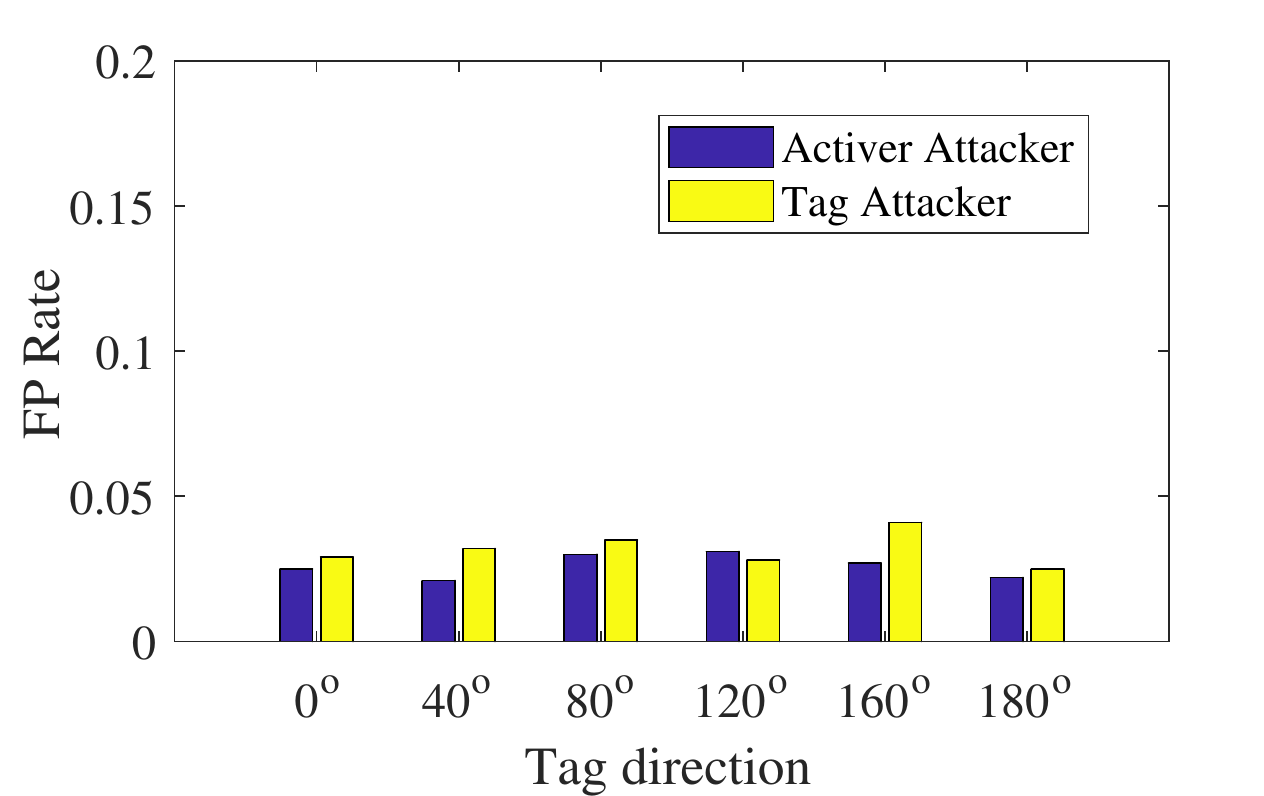}}
	\subfigure[TP rates.]
	{\includegraphics[width=2.7in]{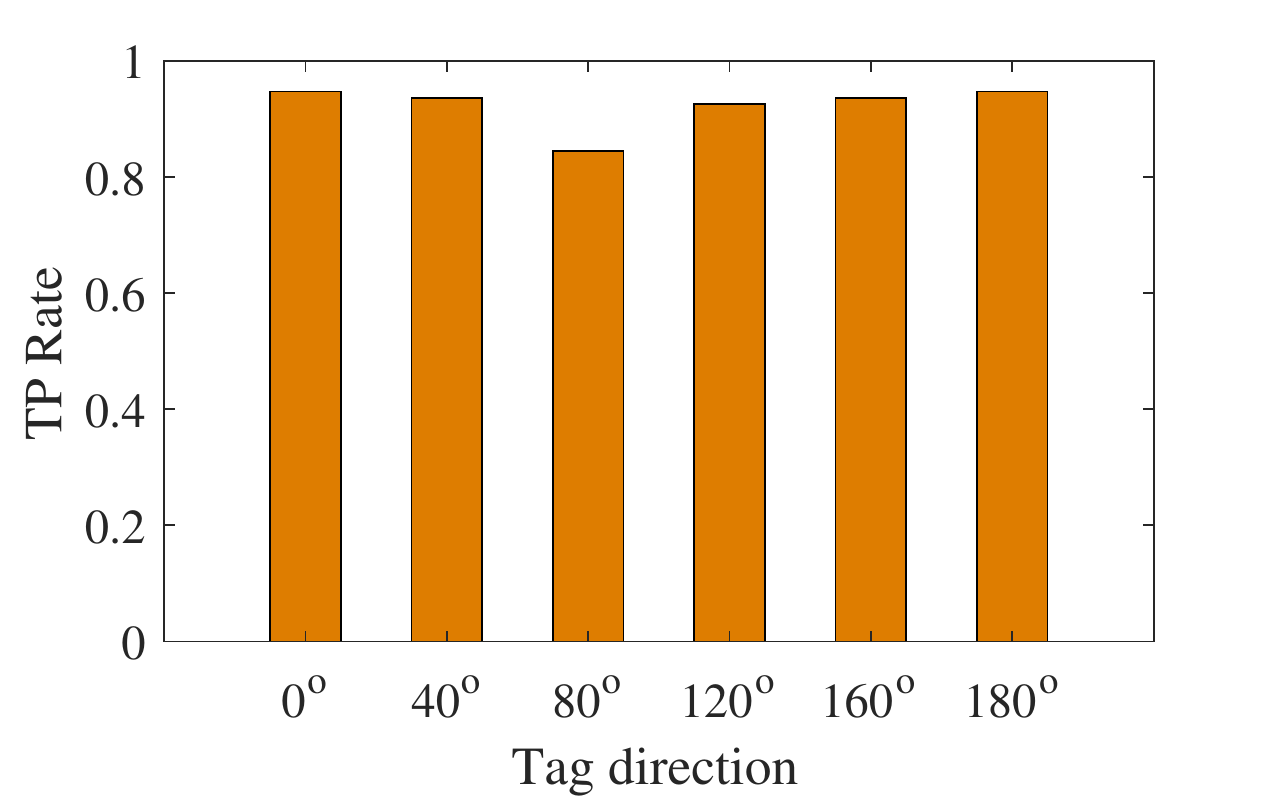}}
	%    \subfigure[Off-body RSS after fluctuation removal.]
	% {\hspace{0.35cm}\includegraphics[width=1.65in]{figs//new/ica_plot/off_removal}}
	\caption{TP and FP rates at different directions of the tag.}
	\label{fig:direction}\vspace{-0.3cm}
\end{figure}

\subsection{Non-Continuous Traffic}\label{sec:non-con} 
In this set of experiments, we evaluate the performance of SecureScatter when the transmitter sends non-continuous packets. In particular, the transmitter conforms to the BLE protocol to send packets at different transmission rates ranging from 10 pkt/s to 320 pkt/s. Then, we use an on-body backscatter tag with a birate of 10~kbps to reflect these non-continuous traffic. Finally, the receiver removes the non-continuous components and perform our system to authenticate on-body tag.

Fig.~\ref{fig:non-con} shows the results. When the transmitter sends packets in a low rate, such as 10 pkt/s, we can only achieve a TP rate of 78.9\%. As expected, the TP rates increase with the rate of the packets transmission. In our experiment, when the packets transmission is large than 40 pkt/s, we can achieve the average rate large than 90\%. Besides, we can still achieve the FP rate lower than 4\%. It means that even in non-continuous traffic, our system can still work well when the rates of the packets transmission are larger than 40 pkt/s.

\subsection{Tag Direction}\label{sec:direction} 
Usually, the antenna of on-body tag can be placed at different directions. In our system, the user authenticates the tag by keeping the tag and receiver static and just moving the transmitter. Thus, the direction of the transmitter is vary all the time. In order to evaluate the performance of the direction of the tag, we control the antennas between the tag and receiver at different angles and then we initialize authentication. 

As illustrated in Fig.~\ref{fig:direction}, when placing the antenna of the tag in different angles ranging from \ang{0} to \ang{180}, we can achieve an average TP rate of 93\% and FP rate of 5\%. It is notable that if the angle is \ang {80}, we can achieve a TP rate of lower than 85\%. That is because if the antennas of the tag and receiver are close to the vertical position, the signal strength will be significantly attenuated by the antenna polarization.

\section{Discussion}\label{sec:discuss}
In this section, we discuss some practical issues that have not been directly addressed in our design.

\subsection{Different Backscatter Technologies} We validate our design using the most widely-adopted backscatter technology that reflects the source signal by changing the antenna impedance. This type of backscatter has been widely used in many applications~\cite{liu2013ambient,huang2018toward,widescatter,nicscatter}. Some recent advances develop new backscatter technologies by shifting phases or frequencies to create backscatter signals \cite{bharadia2015backfi}. Although our detailed algorithm tailored to the basic backscatter cannot be directly applied to these backscatter technologies, the design rationale that exploits on-body backscatter propagation still stands. We believe these on-body propagation signatures can be leveraged to develop proper variants of SecureScatter to authenticate different types of backscatter.

\subsection{Authentication Using One Device} In our experiments, we adopt the same on-body setup as envisioned in the state-of-the-art on-body backscatter~\cite{interscatter}, in which one smart device as the signal source and the other as the receiver. The ideal case is that the signal source and the receiver are incorporated into one on-body smart devices. It requires the device to work in the full duplex mode. In this case, since both the source path and the reflected path still traverse the body surface, our design would also work under this topology.

\subsection{Antenna Types} In our experiments, we use dipole antennas to authenticate the on-body tag. There are many different types of antennas such as patch and inverted-F antennas. The polarization modes of these antennas contain linear, circular and elliptical polarization. Due to the low-cost and form-factor characteristics, linear polarized antenna is widely used for commercial smartphones and wearables. We implement our system using the linear polarized antennas to maximize its wide acceptance.
\section{Related Work} \label{sec:related}%3.9 0.5pp

\textbf{Backscatter communication.} Ambient backscatter originates from RFID systems, which use RFID readers to provide power so as to communicate with battery-free tags~\cite{wang2012efficient}. The differences between them are that backscatter can harvest ambient RF signal and enables two RF-powered devices to communicate by scattering the ambient signal~\cite{liu2013ambient}. In order to enable a network of RF-powered devices to communicate with each other, WiFi backscatter was proposed ~\cite{kellogg2014wi}. Recent innovations in backscatter communication range from the communication distance that backscatter communication can achieve, to the application for wearables. InterScatter \cite{interscatter} bridges the communication between the on-body tags and wearables. There studies pioneer the way to enable backscatter communications in different scenarios, while this paper exploits the security dimension.

\textbf{Authentication.} Conventional authentication schemes adopt arbitrary passwords ~\cite{gehrmann2004manual},~\cite{password} to authenticate smart devices with input interfaces, or exploit certain buttons on devices for association ~\cite{soriente2007beda}. Different from these devices, IoT devices such as on-body tags or implanted devices, lack such interfaces and buttons. 

Proximity-based mechanisms using the radio propagations are close to our system. Proximate~\cite{mathur2011proximate} securely pair two devices in proximity within a half-wavelength distance by comparing their RSS variations. Wanda~\cite{pierson2016wanda} employs two antennas to authenticate the devices in proximity according to the large RSS variation between the two antennas. Similarly, Liang et al. ~\cite{cai2011good} devise a proximity-based secure pairing mechanism using multiple antennas. In our system, instead of leveraging multiple antennas, we exploit the distinct on-body propagation signatures to authenticate the backscatter tag.

Instead of exploiting wireless signals, the data of sensors such as accelerometers and gyroscopes are leveraged to assist security mechanisms~\cite{accelerometers,mayrhofer2009shake} for assistance. In~\cite{mayrhofer2009shake}, the authors consider generating a cryptographic key by shaking the pairing devices and matching features extracted from the sensor data. Besides, some studies exploit the out of band channel, such as ~\cite{schurmann2013secure} and ultrasound~\cite{mayrhofer2007security}, to secure the devices. Ahmed et al. ~\cite{ahmed2015checksum} leverages continuous gestures to authenticate a secure communication channel.

Light-weight authentication protocols have been well designed for RFID tags. These protocols~\cite{peris2006m2ap,peris2006lmap,peris2006emap} consider mutually authenticating the RFID tag by adding random numbers to protect the ID series. Then an authorized reader can access the information and update the index-pseudonym and key in each time after authenticating. Departure from these protocols, our authentication mechanism aims to distinguish backscatter tags that physically on the body from any other devices from afar. On the other hand, these protocols aim to identify each tag by ID series without considering whether the tag is on the body. Thus, the scenarios of them are different and it is inapplicable to use these protocols to authenticate whether the tag is on-body or not.

\textbf{On-body propagation.} There have been a lot of studies on the measurements of the on-body radio propagation. In~\cite{alves2011analytical}, the measurement results indicate that the orientations of on-body antennas impact greatly on the path loss and the different polarized components of the body suffer from different levels of attenuation. On the other hand, the empirical studies in~\cite{hu2007measurements}~\cite{shi2013bana} show the differences in on- and off-body. These studies motive our design of authenticating on-body backscatter based on the on-body radio propagation signatures.

%!TEX root=main.tex

\section{Concluding Remarks}\label{sec:conlude}
This paper presents SecureScatter, a system to authenticate on-body backscatter based on the signatures of on-body radio propagation without any hardware changes. The insight of SecureScatter is that the on-body devices are easily affected by the body movements as well as antenna directions. We evaluate SecureScatter under various environments, with the consideration of body motions and different positions of the tag. We conduct our experiments at both 2.4~GHz and 900~MHz, which are the widely used frequency for wearables and IoT devices. The results show that SecureScatter is robust in both static and dynamic environments by achieving an average TP rate of 93.23\% and FP rates of 3.18\%.  

\vspace{0.2cm}

\begin{acks}
    The research was supported in part by the National Science Foundation of China under Grant 61871441, 61502114, 91738202, and 61531011, Major Program of National Natural Science Foundation of Hubei in China with Grant 2016CFA009.  
\end{acks}

% Bibliography
\bibliographystyle{ACM-Reference-Format}
\bibliography{reference.bib}

\end{document}